%% file: main.tex
\pdfoutput=1
\documentclass[12pt,a4paper]{article}

\usepackage{ifthen} 
\newboolean{pdflatex}
\setboolean{pdflatex}{true} 

\newboolean{articletitles}
\setboolean{articletitles}{true} 

\newboolean{uprightparticles}
\setboolean{uprightparticles}{false} 

\newboolean{inbibliography}
\setboolean{inbibliography}{false} 

\def\paperauthors{LHCb collaboration} 
\def\paperasciititle{Evidence for the decay Bs -> K*bar mu mu} 
\def\papertitle{Evidence for the decay \decay{\Bs}{\Kstarzb\mumu}} 
\def\paperkeywords{{High Energy Physics}, {LHCb}} 
\def\papercopyright{\the\year\ CERN for the benefit of the LHCb collaboration} 
\def\paperlicence{CC-BY-4.0 licence}
\def\paperlicenceurl{https://creativecommons.org/licenses/by/4.0/}

\input{preamble}
\usepackage{longtable} 

\definecolor{colour:Kmm}{RGB}{102,102,102}
\definecolor{colour:pKmm}{RGB}{0,0,102}
\definecolor{colour:Kstmm}{RGB}{204,204,255}
\definecolor{colour:Kstmm:bkg}{RGB}{204,153,204}
\definecolor{colour:combinatorial}{RGB}{102,153,153}
\definecolor{colour:sig}{RGB}{230,230,230}

\begin{document}

\renewcommand{\thefootnote}{\fnsymbol{footnote}}
\setcounter{footnote}{1}

\input{title-LHCb-PAPER}

\renewcommand{\thefootnote}{\arabic{footnote}}
\setcounter{footnote}{0}


\pagestyle{plain} 
\setcounter{page}{1}
\pagenumbering{arabic}


\input{introduction}

\input{detector}

\input{selection}

\input{fit}

\input{results}

\input{systematics}

\input{summary}

\input{acknowledgements}

\input{appendix}


\clearpage

\addcontentsline{toc}{section}{References}
\setboolean{inbibliography}{true}
\bibliographystyle{LHCb}
\bibliography{main,LHCb-PAPER,LHCb-CONF,LHCb-DP,LHCb-TDR}

\newpage

\input{LHCb_Authorship_flat_05-Feb-2018.tex}

\end{document}

%% file: preamble.tex
\usepackage[top=1in, bottom=1.25in, left=1in, right=1in]{geometry}

\usepackage{microtype}
\usepackage{lineno}  
\usepackage{xspace} 
\usepackage{caption} 

\usepackage{graphicx}  
\usepackage{color}
\usepackage{colortbl}
\graphicspath{{./figs/}} 

\usepackage{amsmath} 
\usepackage{amssymb}
\usepackage{amsfonts}
\usepackage{upgreek} 

\newcommand*\patchAmsMathEnvironmentForLineno[1]{%
\expandafter\let\csname old#1\expandafter\endcsname\csname #1\endcsname
\expandafter\let\csname oldend#1\expandafter\endcsname\csname
end#1\endcsname
 \renewenvironment{#1}%
   {\linenomath\csname old#1\endcsname}%
   {\csname oldend#1\endcsname\endlinenomath}%
}
\newcommand*\patchBothAmsMathEnvironmentsForLineno[1]{%
  \patchAmsMathEnvironmentForLineno{#1}%
  \patchAmsMathEnvironmentForLineno{#1*}%
}
\AtBeginDocument{%
\patchBothAmsMathEnvironmentsForLineno{equation}%
\patchBothAmsMathEnvironmentsForLineno{align}%
\patchBothAmsMathEnvironmentsForLineno{flalign}%
\patchBothAmsMathEnvironmentsForLineno{alignat}%
\patchBothAmsMathEnvironmentsForLineno{gather}%
\patchBothAmsMathEnvironmentsForLineno{multline}%
\patchBothAmsMathEnvironmentsForLineno{eqnarray}%
}

\usepackage{hyperxmp}

\usepackage[pdftex,
            pdfauthor={\paperauthors},
            pdftitle={\paperasciititle},
            pdfkeywords={\paperkeywords},
            pdfcopyright={Copyright (C) \papercopyright},
            pdflicenseurl={\paperlicenceurl}]{hyperref}

\usepackage[all]{hypcap} 

\input{lhcb-symbols-def} 

\usepackage{cite} 
\usepackage{mciteplus}

%% file: lhcb-symbols-def.tex
\usepackage{xspace} 
\usepackage{upgreek}

\def\lhcb {\mbox{LHCb}\xspace}

\def\babar  {\mbox{BaBar}\xspace}
\def\belle  {\mbox{Belle}\xspace}

\def\cdf    {\mbox{CDF}\xspace}

\def\MagUp {\mbox{\em Mag\kern -0.05em Up}\xspace}

\ifthenelse{\boolean{uprightparticles}}%
{

 \def\Pmu         {\ensuremath{\upmu}\xspace}

 \def\Ppi         {\ensuremath{\uppi}\xspace}

 \def\Ppsi        {\ensuremath{\uppsi}\xspace}

 \def\PDelta      {\ensuremath{\Delta}\xspace}                 
 \def\PXi      {\ensuremath{\Xi}\xspace}                 
 \def\PLambda      {\ensuremath{\Lambda}\xspace}                 
 \def\PSigma      {\ensuremath{\Sigma}\xspace}                 
 \def\POmega      {\ensuremath{\Omega}\xspace}                 
 \def\PUpsilon      {\ensuremath{\Upsilon}\xspace}                 
 
 \def\PB      {\ensuremath{\mathrm{B}}\xspace}                 
                  
 \def\PD      {\ensuremath{\mathrm{D}}\xspace}

 \def\PJ      {\ensuremath{\mathrm{J}}\xspace}                 
 \def\PK      {\ensuremath{\mathrm{K}}\xspace}

 \def\Pb      {\ensuremath{\mathrm{b}}\xspace}                 
 \def\Pc      {\ensuremath{\mathrm{c}}\xspace}                 
 \def\Pd      {\ensuremath{\mathrm{d}}\xspace}

 \def\Pi      {\ensuremath{\mathrm{i}}\xspace}

 \def\Pp      {\ensuremath{\mathrm{p}}\xspace}

 \def\Ps      {\ensuremath{\mathrm{s}}\xspace}

}
{

 \def\Pmu         {\ensuremath{\mu}\xspace}

 \def\Ppi         {\ensuremath{\pi}\xspace}

 \def\Ppsi        {\ensuremath{\psi}\xspace}                 
                  
 \mathchardef\PDelta="7101
 \mathchardef\PXi="7104
 \mathchardef\PLambda="7103
 \mathchardef\PSigma="7106
 \mathchardef\POmega="710A
 \mathchardef\PUpsilon="7107
                  
 \def\PB      {\ensuremath{B}\xspace}                 
                  
 \def\PD      {\ensuremath{D}\xspace}

 \def\PJ      {\ensuremath{J}\xspace}                 
 \def\PK      {\ensuremath{K}\xspace}

 \def\Pb      {\ensuremath{b}\xspace}                 
 \def\Pc      {\ensuremath{c}\xspace}                 
 \def\Pd      {\ensuremath{d}\xspace}

 \def\Pi      {\ensuremath{i}\xspace}

 \def\Pp      {\ensuremath{p}\xspace}

 \def\Ps      {\ensuremath{s}\xspace}

}

\makeatletter
\ifcase \@ptsize \relax
  \newcommand{\miniscule}{\@setfontsize\miniscule{4}{5}}
\or
  \newcommand{\miniscule}{\@setfontsize\miniscule{5}{6}}
\or
  \newcommand{\miniscule}{\@setfontsize\miniscule{5}{6}}
\fi
\makeatother

\DeclareRobustCommand{\optbar}[1]{\shortstack{{\miniscule (\rule[.5ex]{1.25em}{.18mm})}
  \\ [-.7ex] $#1$}}

\def\mup        {{\ensuremath{\Pmu^+}}\xspace}
\def\mun        {{\ensuremath{\Pmu^-}}\xspace} 
\def\mumu       {{\ensuremath{\Pmu^+\Pmu^-}}\xspace}

\def\ellell     {\ensuremath{\ell^+ \ell^-}\xspace}

\def\dquark    {{\ensuremath{\Pd}}\xspace}

\def\squark    {{\ensuremath{\Ps}}\xspace}

\def\cquark    {{\ensuremath{\Pc}}\xspace}

\def\bquark    {{\ensuremath{\Pb}}\xspace}

\def\pion   {{\ensuremath{\Ppi}}\xspace}

\def\pip    {{\ensuremath{\pion^+}}\xspace}
\def\pim    {{\ensuremath{\pion^-}}\xspace}

\def\kaon    {{\ensuremath{\PK}}\xspace}
  \def\Kbar    {{\kern 0.2em\overline{\kern -0.2em \PK}{}}\xspace}

\def\KorKbar    {\kern 0.18em\optbar{\kern -0.18em K}{}\xspace}

\def\Kp      {{\ensuremath{\kaon^+}}\xspace}
\def\Km      {{\ensuremath{\kaon^-}}\xspace}

\def\Kstarz  {{\ensuremath{\kaon^{*0}}}\xspace}
\def\Kstarzb {{\ensuremath{\Kbar{}^{*0}}}\xspace}
\def\Kstar   {{\ensuremath{\kaon^*}}\xspace}
\def\Kstarb  {{\ensuremath{\Kbar{}^*}}\xspace}

  \def\Dbar    {{\kern 0.2em\overline{\kern -0.2em \PD}{}}\xspace}

\def\DorDbar    {\kern 0.18em\optbar{\kern -0.18em D}{}\xspace}

\def\B       {{\ensuremath{\PB}}\xspace}
\def\Bbar    {{\ensuremath{\kern 0.18em\overline{\kern -0.18em \PB}{}}}\xspace}

\def\BorBbar    {\kern 0.18em\optbar{\kern -0.18em B}{}\xspace}
\def\Bz      {{\ensuremath{\B^0}}\xspace}
\def\Bzb     {{\ensuremath{\Bbar{}^0}}\xspace}
\def\Bu      {{\ensuremath{\B^+}}\xspace}
\def\Bub     {{\ensuremath{\B^-}}\xspace}
\def\Bp      {{\ensuremath{\Bu}}\xspace}
\def\Bm      {{\ensuremath{\Bub}}\xspace}

\def\Bs      {{\ensuremath{\B^0_\squark}}\xspace}

\def\jpsi     {{\ensuremath{{\PJ\mskip -3mu/\mskip -2mu\Ppsi\mskip 2mu}}}\xspace}
\def\psitwos  {{\ensuremath{\Ppsi{(2S)}}}\xspace}

  \def\Y#1S{\ensuremath{\PUpsilon{(#1S)}}\xspace}

\def\proton      {{\ensuremath{\Pp}}\xspace}

\def\Lz          {{\ensuremath{\PLambda}}\xspace}
\def\Lbar        {{\ensuremath{\kern 0.1em\overline{\kern -0.1em\PLambda}}}\xspace}
\def\LorLbar    {\kern 0.18em\optbar{\kern -0.18em \PLambda}{}\xspace}

\def\Lb      {{\ensuremath{\Lz^0_\bquark}}\xspace}

\def\BF         {{\ensuremath{\mathcal{B}}}\xspace}

\newcommand{\decay}[2]{\ensuremath{#1\!\to #2}\xspace}         

\def\to                 {\ensuremath{\rightarrow}\xspace}

\def\qsq       {{\ensuremath{q^2}}\xspace}

\def\AT#1     {\ensuremath{A_{\mathrm{T}}^{#1}}\xspace}           

\def\C#1      {\ensuremath{\mathcal{C}_{#1}}\xspace}                       
\def\Cp#1     {\ensuremath{\mathcal{C}_{#1}^{'}}\xspace}                    
\def\Ceff#1   {\ensuremath{\mathcal{C}_{#1}^{\mathrm{(eff)}}}\xspace}        
\def\Cpeff#1  {\ensuremath{\mathcal{C}_{#1}^{'\mathrm{(eff)}}}\xspace}       
\def\Ope#1    {\ensuremath{\mathcal{O}_{#1}}\xspace}                       
\def\Opep#1   {\ensuremath{\mathcal{O}_{#1}^{'}}\xspace}                    

\newcommand{\tev}{\ifthenelse{\boolean{inbibliography}}{\ensuremath{~T\kern -0.05em eV}}{\ensuremath{\mathrm{\,Te\kern -0.1em V}}}\xspace}
\newcommand{\gev}{\ensuremath{\mathrm{\,Ge\kern -0.1em V}}\xspace}
\newcommand{\mev}{\ensuremath{\mathrm{\,Me\kern -0.1em V}}\xspace}
\newcommand{\kev}{\ensuremath{\mathrm{\,ke\kern -0.1em V}}\xspace}
\newcommand{\ev}{\ensuremath{\mathrm{\,e\kern -0.1em V}}\xspace}
\newcommand{\gevc}{\ensuremath{{\mathrm{\,Ge\kern -0.1em V\!/}c}}\xspace}
\newcommand{\mevc}{\ensuremath{{\mathrm{\,Me\kern -0.1em V\!/}c}}\xspace}
\newcommand{\gevcc}{\ensuremath{{\mathrm{\,Ge\kern -0.1em V\!/}c^2}}\xspace}
\newcommand{\gevgevcccc}{\ensuremath{{\mathrm{\,Ge\kern -0.1em V^2\!/}c^4}}\xspace}
\newcommand{\mevcc}{\ensuremath{{\mathrm{\,Me\kern -0.1em V\!/}c^2}}\xspace}

\def\mum  {\ensuremath{{\,\upmu\mathrm{m}}}\xspace}

\def\invfb   {\ensuremath{\mbox{\,fb}^{-1}}\xspace}

\def\gsim{{~\raise.15em\hbox{$>$}\kern-.85em
          \lower.35em\hbox{$\sim$}~}\xspace}
\def\lsim{{~\raise.15em\hbox{$<$}\kern-.85em
          \lower.35em\hbox{$\sim$}~}\xspace}

\def\ptot       {\mbox{$p$}\xspace}
\def\pt         {\mbox{$p_{\mathrm{ T}}$}\xspace}

\def\mrad{\ensuremath{\mathrm{ \,mrad}}\xspace}

\def\evtgen     {\mbox{\textsc{EvtGen}}\xspace}

\def\geant      {\mbox{\textsc{Geant4}}\xspace}

\def\photos     {\mbox{\textsc{Photos}}\xspace}

\def\pythia     {\mbox{\textsc{Pythia}}\xspace}

\def\tell1  {TELL1\xspace}
\def\ukl1   {UKL1\xspace}



%% file: title-LHCb-PAPER.tex

\begin{titlepage}
\pagenumbering{roman}

\vspace*{-1.5cm}
\centerline{\large EUROPEAN ORGANIZATION FOR NUCLEAR RESEARCH (CERN)}
\vspace*{1.5cm}
\noindent
\begin{tabular*}{\linewidth}{lc@{\extracolsep{\fill}}r@{\extracolsep{0pt}}}
\ifthenelse{\boolean{pdflatex}}
{\vspace*{-1.5cm}\mbox{\!\!\!\includegraphics[width=.14\textwidth]{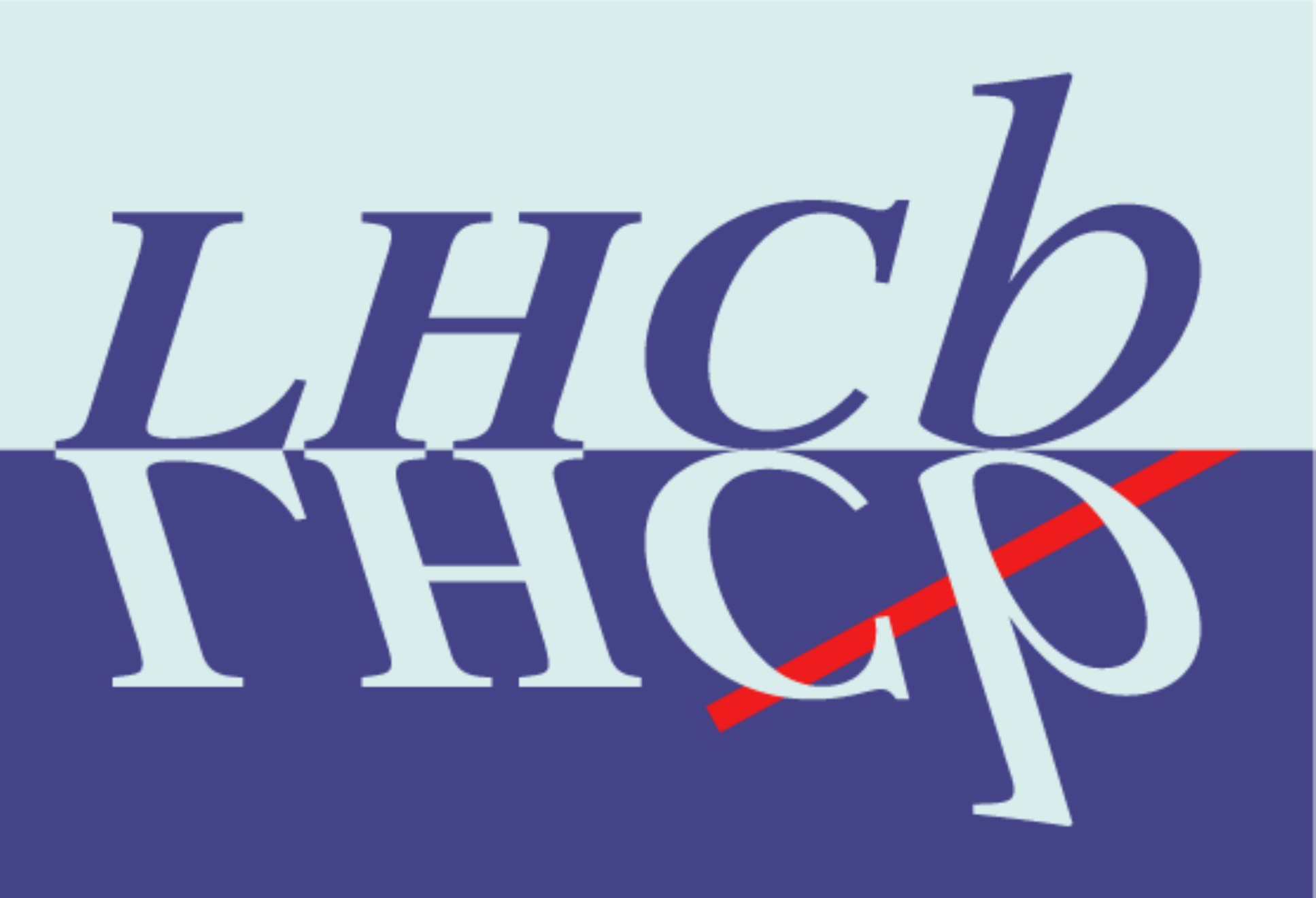}} & &}%
{\vspace*{-1.2cm}\mbox{\!\!\!\includegraphics[width=.12\textwidth]{lhcb-logo.eps}} & &}%
\\
 & & CERN-EP-2018-059 \\  
 & & LHCb-PAPER-2018-004 \\  
 & & 31 July 2018 \\ 
 & & \\
\end{tabular*}

\vspace*{4.0cm}

{\normalfont\bfseries\boldmath\huge
\begin{center}
  \papertitle
\end{center}
}

\vspace*{2.0cm}

\begin{center}
\paperauthors\footnote{Authors are listed at the end of this paper.}
\end{center}

\vspace{\fill}

\begin{abstract}
\noindent
A search for the decay \decay{\Bs}{\Kstarzb\mumu} is presented using data sets corresponding to 1.0, 2.0 and 1.6\invfb of integrated luminosity collected during $pp$ collisions with the LHCb experiment at centre-of-mass energies of 7, 8 and 13\tev, respectively.
An excess is found over the background-only hypothesis with a significance of 3.4 standard deviations.
The branching fraction of the \decay{\Bs}{\Kstarzb\mumu} decay is determined to be
$\BF(\decay{\Bs}{\Kstarzb\mumu}) = [2.9 \pm 1.0\,({\rm stat}) \pm  0.2\,({\rm syst}) \pm 0.3\,({\rm norm})]\times 10^{-8}$,
where the first and second uncertainties are statistical and systematic, respectively.
The third uncertainty is due to limited knowledge of external parameters used to normalise the branching fraction measurement.
\end{abstract}

\vspace*{2.0cm}

\begin{center}
				Published in JHEP 07 (2018) 020
\end{center}

\vspace{\fill}

{\footnotesize
\centerline{\copyright~\papercopyright, \href{\paperlicenceurl}{\paperlicence}.}}
\vspace*{2mm}

\end{titlepage}


\newpage
\setcounter{page}{2}
\mbox{~}
%
%
%
%

\cleardoublepage

%% file: introduction.tex
\section{Introduction}
\label{sec:Introduction}

The decay \decay{\Bs}{\Kstarb(892)^{0}\mumu}, hereafter referred to as \decay{\Bs}{\Kstarzb\mumu},  proceeds via a $\bquark\to\dquark$ flavour-changing neutral-current  (FCNC) transition.
The leading contributions to the amplitude of the decay correspond to loop Feynman diagrams and involve the off-diagonal element $V_{td}$ of the Cabibbo-Kobayashi-Maskawa (CKM) quark-mixing matrix.
This process is consequently rare in the Standard Model of particle physics (SM).
New particles predicted by extensions of the SM can enter in competing diagrams and can significantly enhance or suppress the rate of the decay, see for example Refs.~\cite{Aliev:1998sk,Wang:2007sp}.
Form-factor computations for the $\Bs\to\Kstarzb$ transition have been made using light-cone sum rule~\cite{Straub:2015ica,Ball:2004rg} and lattice QCD~\cite{Horgan:2015vla} techniques.
Standard Model predictions for the branching fraction of the decay are in the range 3--$4 \times 10^{-8}$~\cite{Wu:2006rd,Faustov:2013pca,Kindra:2018ayz}.

The observation of the rare $\bquark\to\dquark\ellell$ FCNC decays \decay{\Bp}{\pip\mumu} and \mbox{\decay{\Lb}{p\pim\mumu}} has been previously reported by the LHCb collaboration in Refs.~\cite{LHCb-PAPER-2015-035} and \cite{LHCb-PAPER-2016-049}, respectively.
Evidence for the decay \decay{\Bz}{\pip\pim\mumu} has also been established in Ref.~\cite{LHCb-PAPER-2014-063}.
The decay \decay{\Bs}{\Kstarzb\mumu} has not yet been observed.
The measured ratio of the \decay{\Bp}{\pip\mumu} and \decay{\Bp}{\Kp\mumu} branching fractions has also been used to determine the ratio of CKM elements $|V_{td}/V_{ts}|$~\cite{Du:2015tda}, exploiting correlations between the $B\to K$ and $B \to \pi$ form-factors in lattice computations.
A similar approach could, in the future, be applied to the ratio of the \decay{\Bs}{\Kstarzb\mumu} and \decay{\Bzb}{\Kstarzb\mumu} decay rates~\cite{Blake:2015tda}.

The decay \decay{\Bzb}{\Kstarzb\mumu}, which involves a $b \to s\ellell$ transition, has been studied extensively by \babar, \belle, \cdf and by the LHC experiments~\cite{Wei:2009zv,Lees:2012tva,Aaltonen:2011cn,CMSKstmm,LHCb-PAPER-2015-051,LHCb-PAPER-2016-012}.
The rate of the decay appears to be systematically lower than current SM predictions.
Global analyses of $b\to s$ processes favour a modification of the SM at the level of 4 to 5 standard deviations~\cite{Altmannshofer:2017fio,Ciuchini:2017mik,Chobanova:2017ghn,Geng:2017svp,Capdevila:2017bsm}.
Similar studies of $b\to d$ processes are important to understand the flavour structure of the underlying theory.

This paper presents a search for the decay \decay{\Bs}{\Kstarzb\mumu}, where the inclusion of charge-conjugate processes is implied throughout, using data collected with the LHCb experiment in $pp$ collisions during Runs~1 and 2 of the LHC.
The data set used in this paper is as follows:
1.0\invfb of integrated luminosity collected at a centre-of-mass energy of 7\tev during Run~1;
2.0\invfb of integrated luminosity collected at a centre-of-mass energy of 8\tev during Run~1;
and 1.6\invfb of integrated luminosity collected at a centre-of-mass energy of 13\tev during Run~2.
Section~\ref{sec:Detector} of this paper describes the LHCb detector and the experimental setup used for the analysis.
Section~\ref{sec:Selection} outlines the selection processes used to identify signal candidates.
Section~\ref{sec:Fit} describes the method used to estimate the number of \decay{\Bs}{\Kstarzb\mumu} decays in the data set.
Section~\ref{sec:Results} describes the determination of the \decay{\Bs}{\Kstarzb\mumu} branching fraction, normalising the number of observed signal decays to the number of \decay{\Bz}{\jpsi\Kstarz} decays present in the data set.
Section~\ref{sec:Systematics} discusses sources of systematic uncertainty on the \decay{\Bs}{\Kstarzb\mumu} branching fraction.
Finally, conclusions are presented in Sec.~\ref{sec:Summary}.

%% file: detector.tex
\section{Detector and simulation}
\label{sec:Detector}

The \lhcb detector~\cite{Alves:2008zz,LHCb-DP-2014-002} is a single-arm forward spectrometer covering the \mbox{pseudorapidity} range $2<\eta <5$,
designed for the study of particles containing \bquark or \cquark quarks.
The detector includes a high-precision tracking system consisting of a silicon-strip vertex detector surrounding the $pp$ interaction region~\cite{LHCb-DP-2014-001}, a large-area silicon-strip detector located upstream of a dipole magnet with a bending power of about $4{\mathrm{\,Tm}}$, and three stations of silicon-strip detectors and straw drift tubes~\cite{LHCb-DP-2013-003} placed downstream of the magnet.
The tracking system provides a measurement of momentum, \ptot, of charged particles with a relative uncertainty that varies from 0.5\% at low momentum to 1.0\% at 200\gevc. The minimum distance of a track to a primary vertex (PV), the impact parameter (IP),  is measured with a resolution of $(15+29/\pt)\mum$, where \pt is the component of the momentum transverse to the beam, in\,\gevc.
Different types of charged hadrons are distinguished using information from two ring-imaging Cherenkov detectors~\cite{LHCb-DP-2012-003}.
Photons, electrons and hadrons are identified by a calorimeter system consisting of scintillating-pad and preshower detectors, an electromagnetic calorimeter and a hadronic calorimeter.
Muons are identified by a system composed of alternating layers of iron and multiwire proportional chambers~\cite{LHCb-DP-2012-002}.

The online event selection is performed by a trigger~\cite{LHCb-DP-2012-004}. 
The trigger consists of a hardware stage, based on information from the calorimeter and muon systems, followed by a software stage, which applies a full event reconstruction.
The signal candidates are required to pass through a hardware trigger that selects events containing at least one muon with $\pt$ greater than 1 to 2\gevc, depending on the data-taking conditions.
The software trigger requires a two-, three- or four-track secondary vertex with a significant displacement from any primary $pp$ interaction vertex.
 At least one charged particle must have a large transverse momentum $\pt > 1\gevc$ and be inconsistent with originating from a PV. 
 A multivariate algorithm~\cite{BBDT} is used for the identification of secondary vertices consistent with the decay of a \bquark hadron.

Samples of simulated \decay{\Bs}{\Kstarzb\mumu}, \decay{\Bzb}{\Kstarzb\mumu}, \decay{\Bs}{\jpsi\Kstarzb} and \mbox{\decay{\Bzb}{\jpsi\Kstarzb}} decays are used to develop an offline event selection and to determine the efficiency to reconstruct the \Bzb and \Bs candidates in the different data-taking periods.
In the simulation, $pp$ collisions are generated using \pythia~\cite{Sjostrand:2007gs} with a specific \lhcb
configuration~\cite{LHCb-PROC-2010-056}.
Decays of hadronic particles are described by \evtgen~\cite{Lange:2001uf}, in which final-state radiation is generated using \photos~\cite{Golonka:2005pn}.
The interaction of the generated particles with the detector, and its response, are implemented using the \geant toolkit~\cite{Allison:2006ve, *Agostinelli:2002hh} as described in Ref.~\cite{LHCb-PROC-2011-006}.
Data-driven corrections are applied to the simulation to account for mismodelling of the detector occupancy and of the $\B_{(s)}^{0}$ meson production kinematics.
The particle identification (PID) performance is measured from data using calibration samples~\cite{LHCb-DP-2014-002}.

%% file: selection.tex
\section{Candidate selection}
\label{sec:Selection}

Signal candidates are formed by combining a \Kstarzb candidate with two oppositely charged tracks, which are identified as muons by the muon system.
The \Kstarzb meson is reconstructed through its decay to the $\Km\pip$ final state with invariant mass within $\pm70\mevcc$ of the known $\Kstar(892)^0$ mass~\cite{PDG2017}.
The muon pair is required to have an invariant mass squared  in the range $0.1 < \qsq < 19.0\gev^{2}/c^{4}$, excluding the region $12.5 < \qsq < 15.0\gev^{2}/c^{4}$ dominated by the \psitwos resonance.
Candidates in the region  $8.0 < \qsq < 11.0\gev^{2}/c^{4}$, which are dominated by decays via a \jpsi resonance, are treated separately in the analysis.
The remaining candidates include \Bs meson decays that produce a dimuon pair through the decay of a light-quark resonance or a charmonium state above the open charm threshold, which are inseparable from the short-distance component of the decay.
These are considered part of the signal in the analysis.

The selection process used in this analysis is similar to that described in Ref.~\cite{LHCb-PAPER-2015-051}.
The four charged tracks are required to each have a significant IP with respect to all PVs in the event and to be consistent with originating from a common vertex.
The $\B_{(s)}^{0}$  meson candidate is required to be consistent with originating from one of the PVs in the event and its decay vertex is required to be well separated from that PV.
The kaon and pion candidates must also be identified as kaon-like and pion-like by a multivariate algorithm~\cite{LHCb-DP-2014-002} based on information from the RICH detectors, tracking system and calorimeters.
The PID requirements are chosen to maximise the sensitivity to a SM-like \decay{\Bs}{\Kstarzb\mumu} signal.

To improve the resolution on the reconstructed $\Km\pip\mumu$ invariant mass, $m(\Km\pip\mumu)$, candidates with an uncertainty larger than $22\mevcc$ on their measured mass are rejected.
The opening angle between every pair of final-state particles is also required to be larger than 5\mrad in the detector.
This requirement removes a possible source of background that arises when the hits associated to a given charged particle are mistakenly used in more than one reconstructed track.
A kinematic fit is also performed, constraining the candidate to originate from its most likely production vertex~\cite{Hulsbergen:2005pu}.
In the kinematic fit of candidates with \qsq in the \jpsi mass window, the dimuon pair is also constrained to the known \jpsi mass.
This mass constraint improves the resolution in $m(\Km\pip\mumu)$ for candidates involving an intermediate \jpsi resonance decay by a factor of two.

Signal candidates are further classified using an artificial neural network~\cite{Feindt:2006pm}.
The neural network is trained using a sample of simulated \decay{\Bzb}{\Kstarzb\mumu} decays as a proxy for the signal decay.
Candidates in data with $m(\Km\pip\mumu) > 5670\mevcc$ are used as a background sample.
This sample is predominantly comprised of combinatorial background, where uncorrelated tracks from the event are mistakenly combined.
The neural network uses the following variables related to the topology of the $\B_{(s)}^{0}$ meson decay:
the angle between the reconstructed momentum vector of the $\B_{(s)}^{0}$ meson and the vector connecting the PV and the decay vertex of the $\B_{(s)}^{0}$ candidate;
the IP, \pt and proper decay time of the $\B_{(s)}^{0}$ candidate;
the vertex fit quality of the $\B_{(s)}^{0}$ decay vertex and of the dimuon pair;
the minimum and maximum \pt of the final-state particles
and for the Run~1 data set a measure of the isolation of the final-state particles in the detector.
It has been verified that the distribution of the variables used as input to, and the output distribution from, the classifier agree between the simulation and the data.
The output of the neural network is transformed such that it is uniform in the range 0--1 on the signal proxy.
Candidates with neural network response below 0.05 are rejected in the subsequent analysis.
This requirement removes a background-dominated part of the data sample.
The neural network response is validated on simulated \decay{\Bzb}{\Kstarzb\mumu} and \decay{\Bs}{\Kstarzb\mumu} decays to ensure that it does not introduce any bias in $m(\Km\pip\mumu)$.

Finally, a number of vetoes are applied to reject specific sources of background.
Signal candidates are rejected if the pion candidate has a nonnegligible probability to be a kaon and if the \Kp\Km invariant mass, after assigning the kaon mass to the pion candidate, is consistent within $10\mevcc$ of the known $\phi(1020)$ meson mass.
This veto removes 98\% of \decay{\Bs}{\phi\mumu} decays inside the $\phi(1020)$ mass window.
Candidates are also rejected if the kaon or pion are identifiable as a muon and the $\Km\mup$ or $\pip\mun$ mass, after assigning the muon mass hypothesis to the kaon or pion candidate, are consistent with that of a \jpsi or \psitwos meson (within $\pm 60\mevcc$ of their known masses).

%% file: fit.tex
\section{Signal yields}
\label{sec:Fit}

In order to maximise sensitivity to a \decay{\Bs}{\Kstarzb\mumu} signal, candidates are divided into regions of neural network response.
The candidates are also divided based on the two data-taking periods, Run~1 and Run~2.
Four regions of neural network response are selected for each data-taking period, each containing an equal amount of expected signal decays.
The yield of the \decay{\Bs}{\Kstarzb\mumu} decay is determined by performing a simultaneous unbinned maximum likelihood fit to the $m(\Km\pip\mumu)$ distribution of the eight resulting subsets of the data.

In the likelihood fit, the signal lineshape of both the \Bzb and the \decay{\Bs}{\Kstarzb\mumu} decays is described by the sum of three functions: a Gaussian function with a power-law tail on the lower-side of its peak, used to describe final-state radiation and energy loss in the detector; a Gaussian function with a power-law tail on the upper-side of its peak, used to describe the non-Gaussian tails of the signal mass distribution at large masses; and an additional Gaussian function to account for differences in the per-candidate resolution of the reconstructed mass.
The two functions with power-law tails share a common width and all three functions share a common peak position.
The \Bs peak position is displaced from that of the \Bz by 87.5\mevcc~\cite{LHCb-PAPER-2011-035}.
The relative fractions of each function are fixed from fits to simulated \Bzb and \decay{\Bs}{\Kstarzb\mumu} decays.
The widths of the functions and all of the tail parameters are also fixed from the simulation, except for an overall scaling of the widths and of the tail parameters to allow for potential data-simulation differences.
The peak position and these scale factors are obtained from a fit to candidates with the dimuon in the \jpsi mass window, where the mass constraint on the dimuon mass has not been applied.
The result of this fit is shown in the appendix in Fig.~\ref{fig:appendix:fits:jpsinocon}.
In the fit to the data, the widths vary from their values in the simulation by 10 to 15\%.
The turn-on point of the upper tail (relative to the width of the distribution) is found to be consistent between data and simulation.

After applying the selection procedure, the background predominately comprises combinatorial background.
The combinatorial background is described in the fit by a separate exponential function in each subset of the data.
A number of other sources of background are accounted for in the fit.
The decay \decay{\Bz}{\Kstarz\mumu} forms a source of background if the  kaon is mistakenly identified as the pion and vice versa.
The shape of this background is taken from the simulation.
The yield of the background is constrained relative to that of the \decay{\Bzb}{\Kstarzb\mumu} decay based on measurements of the kaon-to-pion and pion-to-kaon misidentification probabilities in the PID calibration samples.
The decay \decay{\Lb}{\proton\Km\mumu} forms a source of background if the final-state hadrons are misidentified.
This background is constrained from a control region in the data, by modifying the PID requirements on the candidates to preferentially select $p\Km$ rather than $\Km\pip$ combinations.
The shape of this background is modelled in the fit by Crystal Ball functions.
The yield in each subset of the data is constrained using the proton and kaon identification and misidentification probabilities determined from the PID calibration samples.
The decay \decay{\Bm}{\Km\mumu} forms a source of background if a pion from the event is mistakenly combined with the particles coming from the \Bm meson decay.
The background contribution from \decay{\Bm}{\Km\mumu}  decays is determined from a control region in the data, by selecting candidates with a $\Km\mumu$ invariant mass that is consistent with the known \Bm mass.
This background is only visible for the candidates with \qsq in the \jpsi mass region.
The shape of the background in the fit is modelled by Crystal Ball functions.
Several other sources of background are considered but are found to have a negligible contribution to the fit.
These sources include semileptonic decays of \bquark hadrons via intermediate open-charm states and fully hadronic \bquark-hadron decays.
The background from semileptonic decays is predominantly reconstructed at low $m(\Km\pip\mumu)$ and does not contribute to the analysis.
Fully hadronic \bquark-hadron decays contribute at the level of 1 to 2 candidates at masses close to the known \Bs mass.
This background is neglected in the analysis but is considered as a source of systematic uncertainty in Sec.~\ref{sec:Systematics}.

\begin{figure}[!t]
\centering
\includegraphics[width=0.49\linewidth]{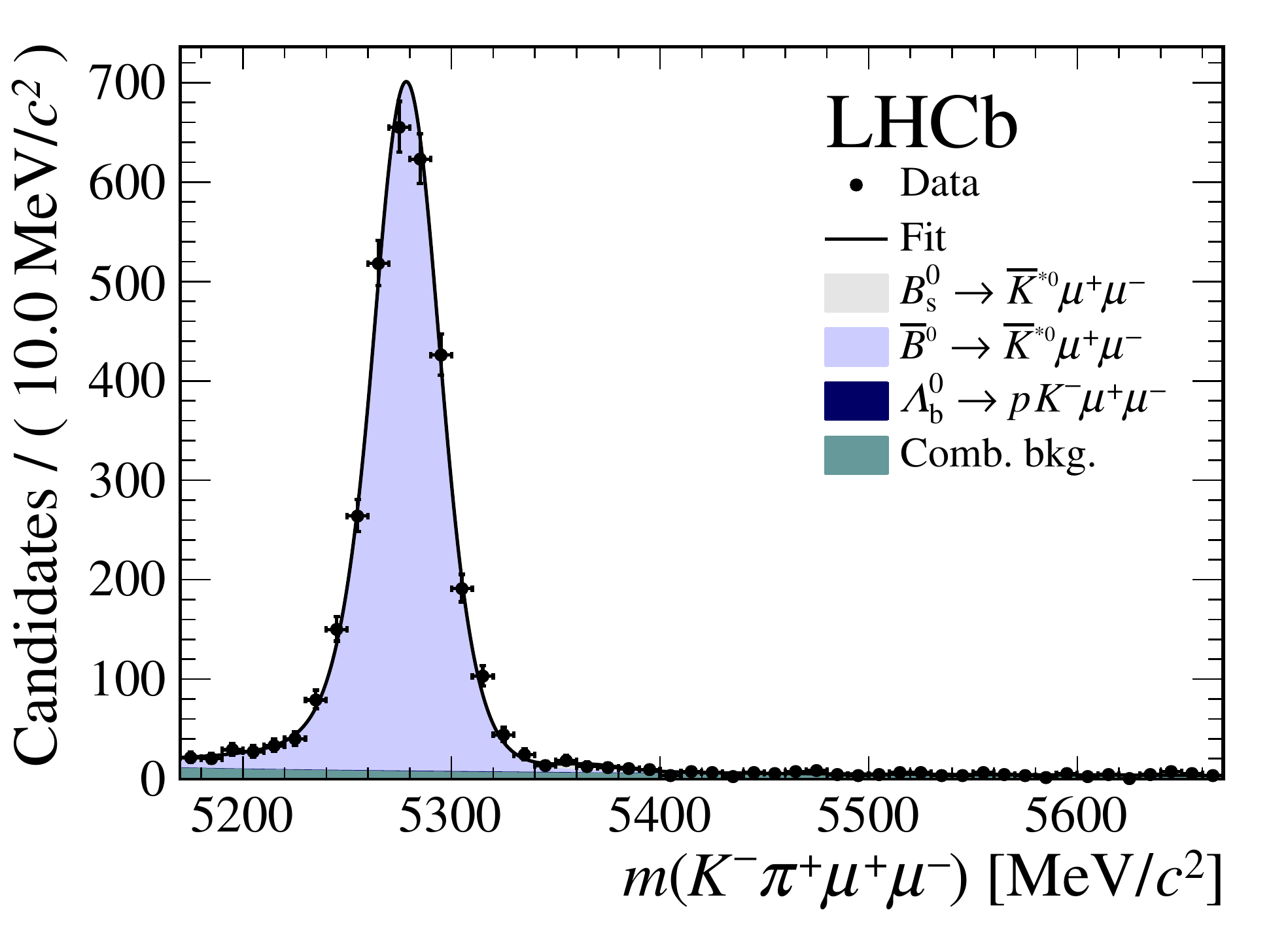}
\includegraphics[width=0.49\linewidth]{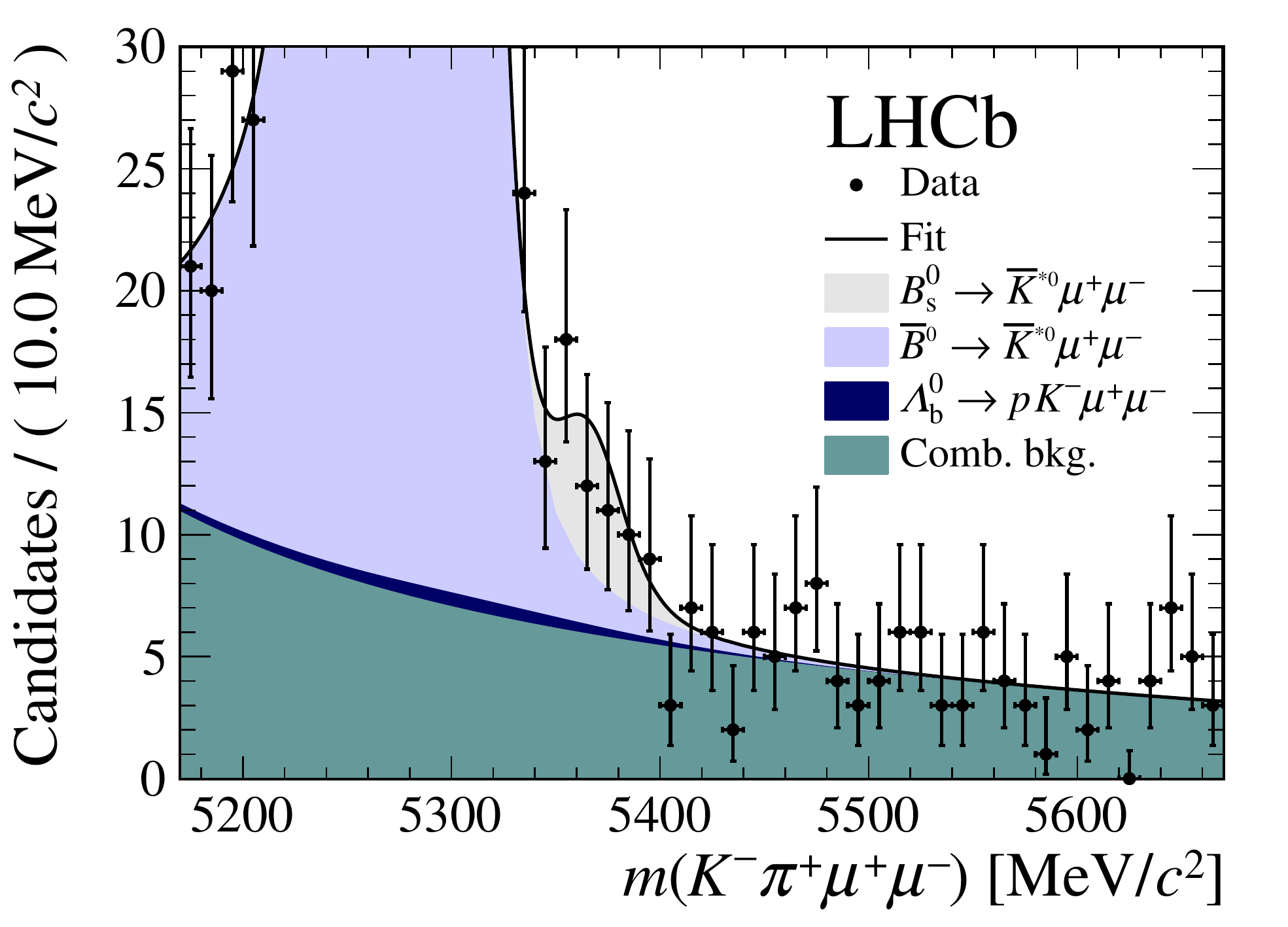}\\
\caption{
Distribution of reconstructed $\Km\pip\mumu$ invariant mass of candidates outside the \jpsi and \psitwos mass regions, summing the three highest neural network response bins of each run condition.
The candidates are shown (left) over the full range and (right) over a restricted vertical range to emphasise the \decay{\Bs}{\Kstarzb\mumu} component.
The solid line indicates a combination of the results of the fits to the individual bins.
Components are detailed in the legend, where they are shown in the same order as they are stacked in the figure.
The background from misidentified \decay{\Bz}{\Kstarz\mumu} decays is included in the \decay{\Bzb}{\Kstarzb\mumu} component.
}
\label{fig:fits:mumu}
\end{figure}

\begin{figure}[!t]
\centering
\includegraphics[width=0.49\linewidth]{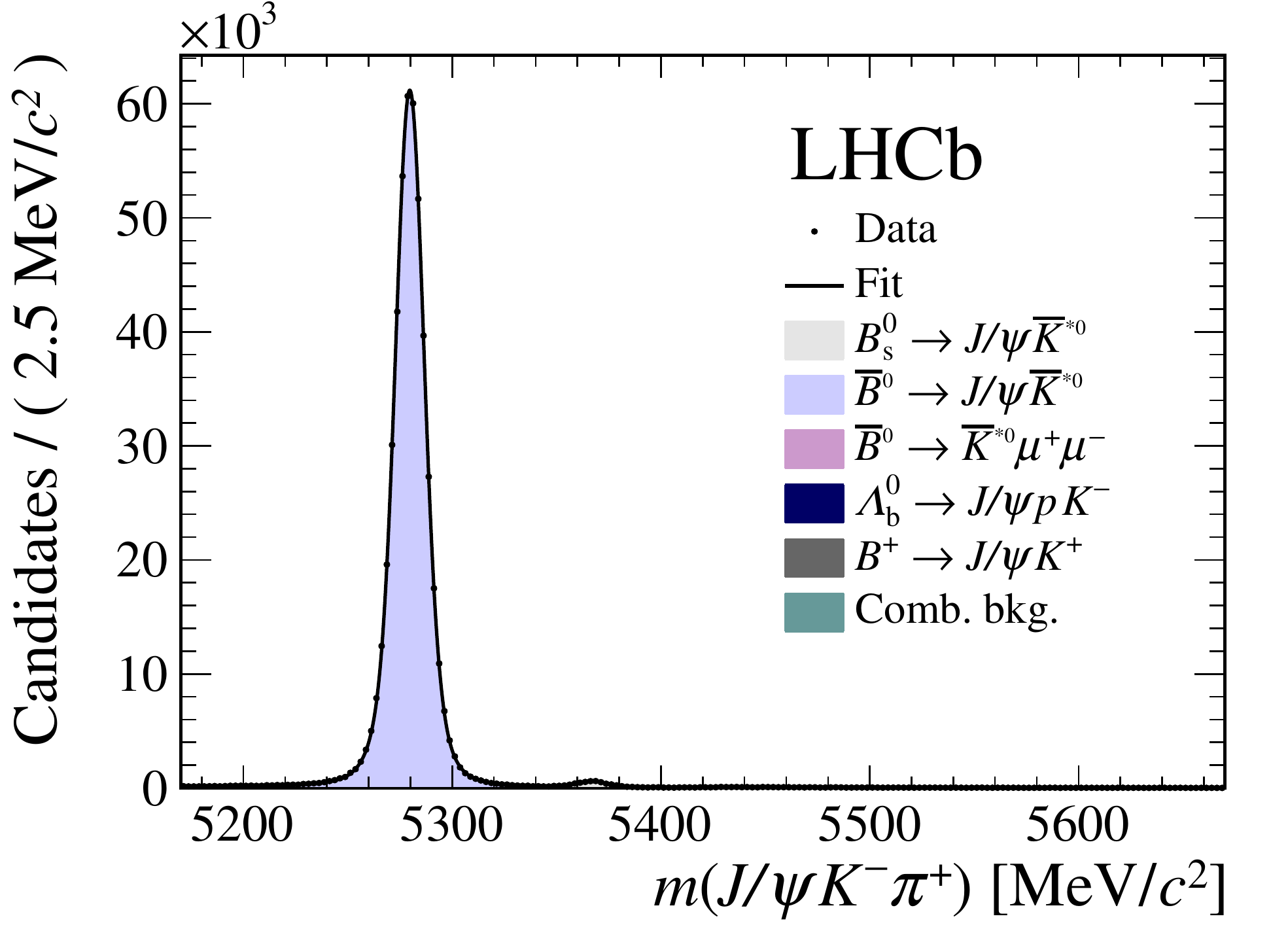}
\includegraphics[width=0.49\linewidth]{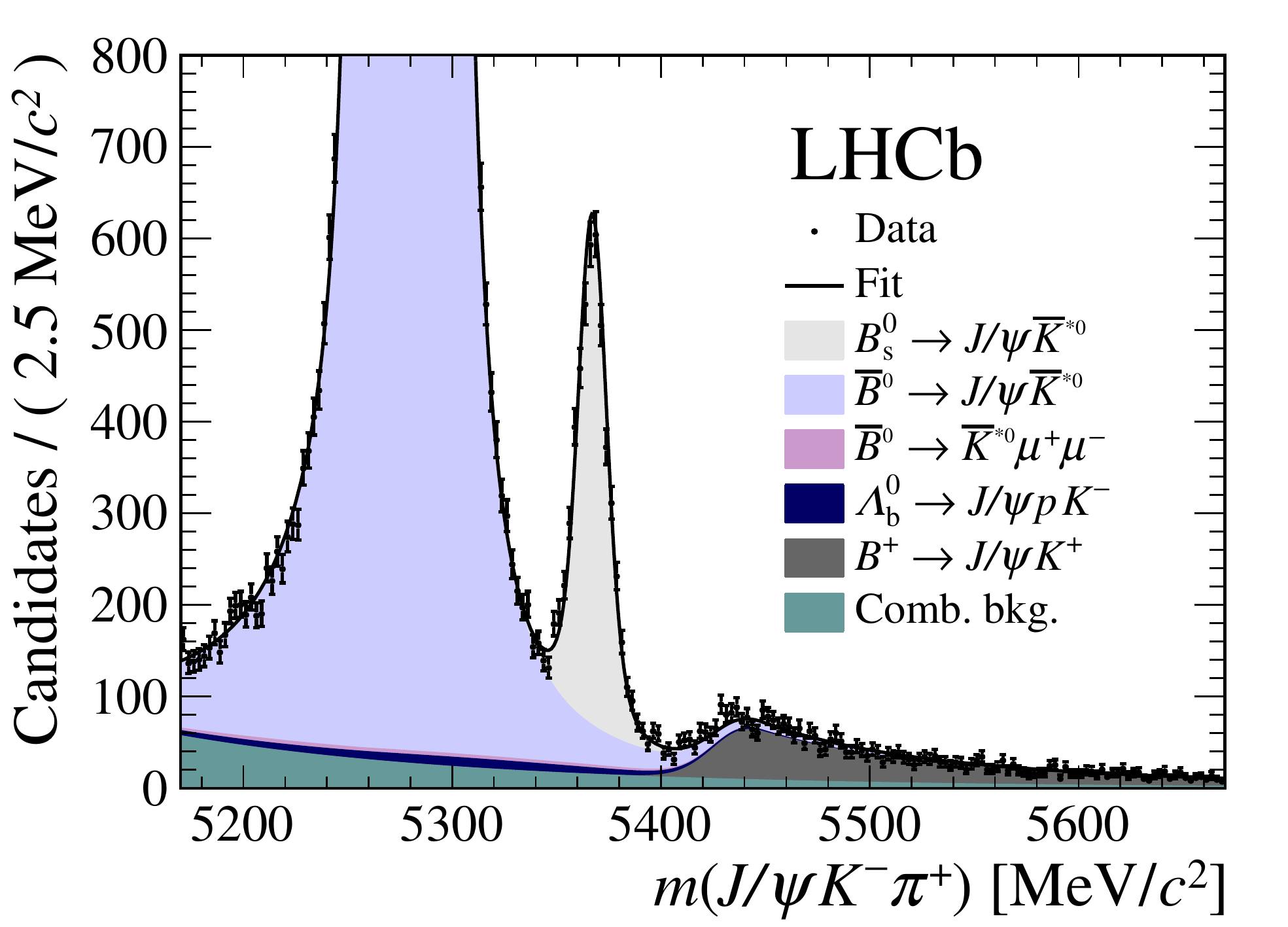}\\
\caption{
Distribution of reconstructed $\jpsi\Km\pip$ invariant mass of the candidates in the \jpsi mass region summing the three highest neural network response bins of each run condition, shown (left) over the full range and (right) over a restricted vertical range to emphasise the \decay{\Bs}{\jpsi\Kstarzb} component.
The solid line indicates a combination of the results of the fits to the individual bins.
Components are detailed in the legend, where they are shown in the same order as they are stacked in the figure.
The background from misidentified \decay{\Bz}{\jpsi\Kstarz} decays is included in the \decay{\Bzb}{\jpsi\Kstarzb} component.
}
\label{fig:fits:jpsi}
\end{figure}

\begin{figure}[!t]
\centering
\includegraphics[width=0.65\linewidth]{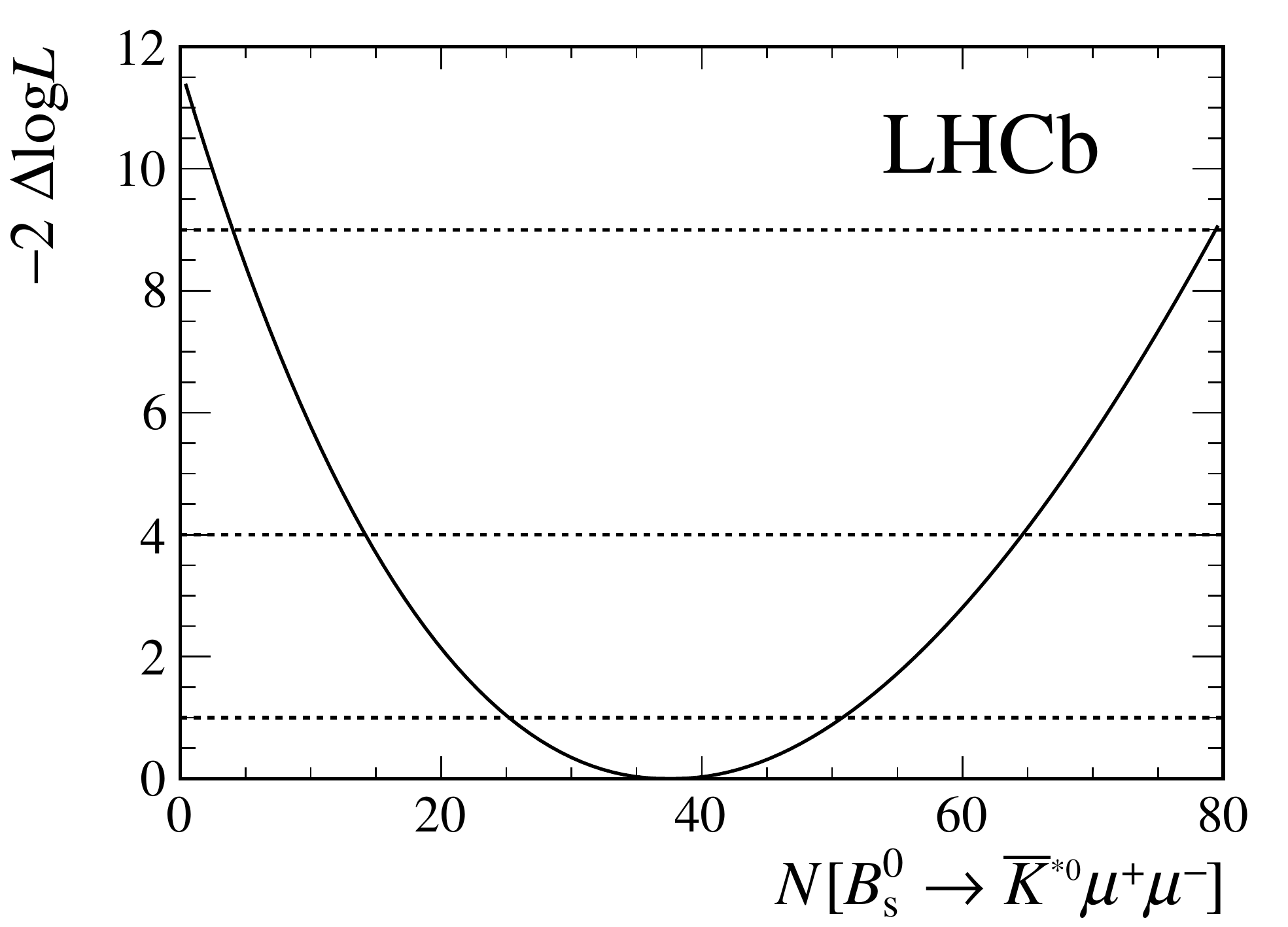}
\caption{
Change in log-likelihood from the simultaneous fit to the candidates in the two data-taking periods and the different bins of neural network response, as a function of the \decay{\Bs}{\Kstarzb\mumu} yield.
Systematic uncertainties on the yield have been included in the likelihood.
}
\label{fig:fits:loglikelihood}
\end{figure}

Figure~\ref{fig:fits:mumu} shows the fit to the candidates, where the result of the fit in the three most signal-like neural network response bins for each data-taking period has been combined.
Candidates in the least signal-like bin are not included.
This bin has a much higher level of combinatorial background and would visually obscure any \Bs signal.
The dominant contribution in the fit is the \decay{\Bzb}{\Kstarzb\mumu} decay.
Figure~\ref{fig:fits:jpsi} shows the fit to the mass-constrained candidates in the \jpsi mass region, also with the three highest neural network response bins for each data taking period combined.
In this fit,  a small background component from \decay{\Bzb}{\Kstarzb\mumu} decays is included.
This background has the same final state but is constrained to the wrong dimuon mass and becomes a broad component in the fit.
The fit results in individual bins of neural network response are shown in the appendix in Figs.~\ref{fig:appendix:fits:mumu} and \ref{fig:appendix:fits:jpsi}.
Summing over the bins of neural network response and data-taking periods, the yields are:
$627\,244 \pm 837$ for the \decay{\Bzb}{\jpsi\Kstarzb} decay,
$5730 \pm 94$ for the \decay{\Bs}{\jpsi\Kstarzb} decay,
$4157 \pm 72$ for the \decay{\Bzb}{\Kstarzb\mumu} decay, and
$38 \pm 12$ for the \decay{\Bs}{\Kstarzb\mumu} decay.
No correction has been made to these yields to account for cases where the $\Km\pip$ system does not originate from a $\Kstarb(892)^{0}$ decay.
Contamination from non-\Kstarzb decays is discussed further in Sec.~\ref{sec:Results}.
Using Wilks' theorem, and a likelihood ratio test between the signal-plus-background and the background-only hypothesis, the significance of the \decay{\Bs}{\Kstarzb\mumu} yield is determined to be $\sqrt{ -2 \log (L_{S+B}/L_B}) = 3.4$ standard deviations.
The signal significance has been validated using pseudoexperiments generated under the null hypothesis.
This includes the systematic uncertainties on the yield discussed in Sec.~\ref{sec:Systematics}.
Figure~\ref{fig:fits:loglikelihood} shows the variation of the log-likelihood of the simultaneous fit as a function of the \decay{\Bs}{\Kstarzb\mumu} yield.

%% file: results.tex
\section{Results}
\label{sec:Results}

The branching fraction of the \decay{\Bs}{\Kstarzb\mumu} decay is determined with respect to that of \decay{\Bzb}{\jpsi\Kstarzb} according to
\begin{align}
\begin{split}
\hspace{-20pt}
\BF(\decay{\Bs}{\Kstarzb\mumu}) =
\BF(&\decay{\Bzb}{\jpsi\Kstarzb}) \BF(\decay{\jpsi}{\mumu}) \\
& \times \frac{f_d}{f_s} \frac{N(\Bs\to\Kstarzb\mumu)}{\varepsilon(\Bs\to\Kstarzb\mumu)}\frac{\varepsilon(\Bzb\to\jpsi\Kstarzb)}{N(\Bzb\to\jpsi\Kstarzb)}\,.
\end{split}
\label{eq:result}
\end{align}
Here, $N$ is the yield for a given decay mode determined from the fit to  $m(\Km\pip\mumu)$ or $m(\jpsi\Km\pip)$ and $\varepsilon$ is the efficiency to reconstruct and select the given decay mode.
The ratio $f_s/f_d$ is the relative production fraction of \Bs and \Bzb mesons in $pp$ collisions.

The efficiency to trigger, reconstruct and select each of the decay modes is determined from the simulation after applying the data-driven corrections.
The efficiency for the \decay{\Bs}{\Kstarzb\mumu} decay is corrected to account for events in the vetoed \qsq regions following the same prescription as Ref.~\cite{LHCb-PAPER-2016-012}.
The efficiency corrected yields are further corrected for contamination from decays with the $\Km\pip$ system in an S-wave configuration.
For the decay \decay{\Bs}{\jpsi\Kstarzb}, the S-wave fraction of $F_{\rm S}(\decay{\Bzb}{\jpsi\Kstarzb}) = (6.4\pm 0.3\pm 1.0)\%$ determined in Ref.~\cite{LHCb-PAPER-2013-023} is used.
The S-wave contamination of the \decay{\Bs}{\Kstarzb\mumu} decay is unknown but it is assumed to be at a similar level to that of the \decay{\Bzb}{\Kstarzb\mumu} decay.
The full size of the S-wave correction is taken as a systematic uncertainty.
The S-wave contamination of the \decay{\Bzb}{\Kstarzb\mumu} decay is determined using the model from Ref.~\cite{LHCb-PAPER-2016-012}.
This model predicts an S-wave fraction of $F_{\rm S}(\decay{\Bzb}{\Kstarzb\mumu}) = (3.4\pm0.8)\%$ in the $\Km\pip$ mass window used in this analysis.

The ratio of production fractions, $f_{s}/f_{d}$, has been measured at 7 and 8\tev to be $f_s/f_d = 0.259 \pm 0.015$ in the LHCb detector acceptance~\cite{fsfd}.
The production fraction at 13\tev has been shown to be consistent with that of the 7 and 8\tev data in Ref.~\cite{LHCb-PAPER-2017-001}.
The production fraction at 13\tev has also been validated in this analysis by comparing the efficiency-corrected yields of the \Bzb and the \decay{\Bs}{\jpsi\Kstarzb} decays in bins of the $\B_{(s)}^{0}$ meson \pt.
Taking the branching fractions of the decays \decay{\Bzb}{\jpsi\Kstarzb} and \decay{\jpsi}{\mumu}  to be
$(1.19 \pm 0.01 \pm 0.08) \times 10^{-3}$~\cite{Chilikin:2014bkk} and $(5.96 \pm 0.03 )\%$~\cite{PDG2017}, respectively, results in a branching fraction for the \decay{\Bs}{\Kstarzb\mumu} decay of
\begin{displaymath}
\BF(\decay{\Bs}{\Kstarzb\mumu}) =
[2.9 \pm 1.0\,({\rm stat}) \pm 0.2\,({\rm syst}) \pm 0.3\,({\rm norm})]\times 10^{-8}~.
\end{displaymath}
The first and second uncertainties are statistical and systematic, respectively.
The third uncertainty is due to limited knowledge of the external parameters used to normalise the observed yield.
This comprises the uncertainty on the external branching fraction measurements, on $f_s/f_d$, $F_{\rm S}(\decay{\Bzb}{\jpsi\Kstarzb})$ and $F_{\rm S}(\decay{\Bs}{\Kstarzb\mumu})$.

A measurement of the branching fraction of the \decay{\Bs}{\Kstarzb\mumu} decay relative to that of \decay{\Bs}{\jpsi\Kstarzb} is also made.
The S-wave contamination of the \decay{\Bs}{\jpsi\Kstarzb} decay is corrected for by using the measurements of $F_{\rm S}$ in bins of $m(\Km\pip)$ from Ref.~\cite{LHCb-PAPER-2015-034}, scaled according to the model in Ref.~\cite{LHCb-PAPER-2016-012}, giving $F_{\rm S}(\decay{\Bs}{\jpsi\Kstarzb}) = (16.0\pm 3.0)\%$.
The resulting ratio of branching fractions is
\begin{displaymath}
\frac{\BF(\decay{\Bs}{\Kstarzb\mumu})}{\BF(\decay{\Bs}{\jpsi \Kstarzb})\BF(\decay{\jpsi}{\mumu})} =
[1.4 \pm 0.4\,({\rm stat}) \pm 0.1\,({\rm syst}) \pm 0.1\,({\rm norm})]\times 10^{-2}~,
\end{displaymath}
where the third uncertainty is due to $F_{\rm S}(\decay{\Bs}{\jpsi\Kstarzb})$ and $F_{\rm S}(\decay{\Bs}{\Kstarzb\mumu})$.

In order to determine the ratio $|V_{td}/V_{ts}|$ it is also useful to extract the ratio
\begin{align}
\frac{\BF(\decay{\Bs}{\Kstarzb\mumu})}{\BF(\decay{\Bzb}{\Kstarzb\mumu})} & =  \frac{f_d}{f_s} \frac{N(\Bs\to\Kstarzb\mumu)}{\varepsilon(\Bs\to\Kstarzb\mumu)}\frac{\varepsilon(\Bzb\to\Kstarzb\mumu)}{N(\Bzb\to\Kstarzb\mumu)} \\
& = [3.3 \pm 1.1\,({\rm stat}) \pm 0.3\,({\rm syst}) \pm 0.2\,({\rm norm})]\times 10^{-2}~, \nonumber
\end{align}
where the third uncertainty corresponds to the uncertainties on $f_{s}/f_{d}$, $F_{\rm S}(\decay{\Bzb}{\Kstarzb\mumu})$ and \mbox{$F_{\rm S}(\decay{\Bs}{\Kstarzb\mumu})$}.

%% file: systematics.tex
\section{Systematic uncertainties}
\label{sec:Systematics}

The measurements presented in Sec.~\ref{sec:Results} are performed relative to decays that have the same final-state particles as the \decay{\Bs}{\Kstarzb\mumu} decay.
Consequently, many potential sources of systematic uncertainty largely cancel in the ratios.
The remaining sources of systematic uncertainty are discussed below and are summarised in Table~\ref{tab:syst}.
Only systematic uncertainties that have an effect on the measured yield are considered when evaluating the significance of the observed signal.
These are systematic uncertainties related to the signal resolution, neural network binning scheme and the residual backgrounds at $m(\Km\pip\mumu)$ close to the known \Bs meson mass.

The $m(\Km\pip\mumu)$ model used to describe the decays \Bzb and \decay{\Bs}{\Kstarzb\mumu} is taken from the simulation with a simple scaling of the width and tail parameters based on the fit to the data in the \jpsi mass region.
Any difference in the \qsq spectrum of the simulation and the data could result in a small mismodelling of the lineshape.
To account for this possibility, the width of the $m(\Km\pip\mumu)$ resolution model is allowed to vary within 0.5\mevcc in the fit.
This covers the full variation in the simulation of the width across the allowed \qsq range and contributes 0.1\% to the systematic uncertainty.
A final uncertainty on the signal lineshape is evaluated based on the difference in fits to the candidates in the \jpsi mass region with and without the constraint on the dimuon mass.
A systematic uncertainty of 0.5\% is assigned, taken as the difference in efficiency-corrected \decay{\Bzb}{\jpsi\Kstarzb} yields between these two fits.
In addition, an alternative parameterisation with an exponential tail rather than a power-law tail is tested for the lineshape describing the \Lb background.
The difference in yields between the two models results in a systematic uncertainty of 0.1\% on the \decay{\Bs}{\Kstarzb\mumu} yield.
The total uncertainty related to mass lineshapes is taken as the sum in quadrature of the uncertainties.

\begin{table}[!tb]
\caption{
Main sources of systematic uncertainty considered on the branching fraction measurements.
The first uncertainty applies to the measurement of $\BF(\decay{\Bs}{\Kstarzb\mumu})$,
the second to $\BF(\decay{\Bs}{\Kstarzb\mumu})/\BF(\decay{\Bzb}{\Kstarzb\mumu})$
and the third to $\BF(\decay{\Bs}{\Kstarzb\mumu})/\BF(\decay{\Bs}{\jpsi\Kstarzb})$, respectively.
A description of the different contributions can be found in the text.
The first three sources of uncertainty affect the measured yield of the signal decay.
The total uncertainty is the sum in quadrature of the individual sources.
The final row indicates the additional uncertainty arising from the uncertainties on external parameters used in the measurements.
}
\centering
\begin{tabular}{lccc}
\hline
 & \multicolumn{3}{c}{Uncertainties} \\
Source & ${\scriptstyle \BF(\decay{\Bs}{\Kstarzb\mumu})}$ & $\tfrac{\BF(\decay{\Bs}{\Kstarzb\mumu})}{\BF(\decay{\Bzb}{\Kstarzb\mumu})}$ & $\tfrac{\BF(\decay{\Bs}{\Kstarzb\mumu})}{\BF(\decay{\Bs}{\jpsi\Kstarzb})}$ \\[5pt]
\hline
Mass lineshapes & 0.5\% & 0.5\% & 0.5\%  \\
Neural network response & 0.5\% &  0.5\% &  0.5\% \\
Residual background & 2.0\% & 2.0\% & 2.0\% \\
\hline
Decay models & 4.0\% & 4.0\% & 4.0\%\\
Non-\Kstarzb states & 3.4\% &  3.4\% &  3.4\% \\
Efficiency & 1.3\% & 1.5\% & 1.4\%\\
Data-simulation differences & 2.2\% & 2.2\% & 0.8\% \\
\hline
Total systematic uncertainty & 6.2\% & 6.3\% & 5.9\% \\
\hline
External parameters & 8.9\% & 5.9\% & 4.0\%\\
\hline
\end{tabular}
\label{tab:syst}
\end{table}

The systematic uncertainty related to the relative efficiencies in each neural network response bin is evaluated in two parts:
an uncertainty due to the limited size of the simulation sample used to determine the relative fractions and an uncertainty due to differences between simulated samples and the data.
The latter is evaluated by correcting the fraction of \decay{\Bs}{\Kstarzb\mumu} decays in each neural network response bin by the measured difference between simulation and data for the  \decay{\Bz}{\jpsi\Kstarzb} decays.
The combination of these uncertainties is 0.5\%.

Sources of background from hadronic \bquark-hadron decays, where two of the final-state hadrons are misidentified as muons, are neglected in the final fit to the $\Kstarzb\mumu$ candidates.
These backgrounds are estimated to contribute 1 to 2 candidates at $m(\Km\pip\mumu)$ close to the known \Bs mass.
The resulting systematic uncertainty on the \decay{\Bs}{\Kstarzb\mumu} yield is estimated to be 2\%.
The background is negligible compared to the \Bzb yield.
The background yield from \Lb decays is constrained using PID efficiencies from control samples and these efficiencies have an associated systematic uncertainty.
This uncertainty is accounted for in the statistical uncertainty of the fit and is negligible.

Other sources of systematic uncertainties are associated to the normalisation of the observed yield for the measurements of the branching fraction and branching-fraction ratios.
The largest source of systematic uncertainty on both $\BF(\decay{\Bs}{\Kstarzb\mumu})$ and the branching-fraction ratio measurements is associated to how well external parameters are known: there is a 5.8\% uncertainty on the ratio of the \Bs and \Bzb fragmentation fractions, a 1.1\% systematic uncertainty due to $F_{\rm S}(\mbox{\decay{\Bzb}{\jpsi\Kstarzb}})$, a 0.8\% uncertainty due to $F_{\rm S}(\mbox{\decay{\Bzb}{\Kstarzb\mumu}})$, a 4.0\% uncertainty due to $F_{\rm S}(\mbox{\decay{\Bs}{\jpsi\Kstarzb}})$ and a 6.8\% uncertainty on $\BF(\mbox{\decay{\Bzb}{\jpsi\Kstarzb}})$.
It is assumed that these external uncertainties are uncorrelated.

The second largest source of uncertainty is due to how well the amplitudes for the \decay{\Bzb}{\jpsi\Kstarzb}, \decay{\Bs}{\jpsi\Kstarzb}, \decay{\Bzb}{\Kstarzb\mumu}, and \decay{\Bs}{\Kstarzb\mumu} decays are known.
The uncertainty on the decay structure leads to an uncertainty on the efficiencies used to correct the observed yields.
The amplitude structure of the \decay{\Bzb}{\jpsi\Km\pip} decay has been studied in Refs.~\cite{Chilikin:2014bkk,LHCb-PAPER-2013-023},
and the amplitude structure of the \decay{\Bs}{\jpsi\Km\pip} decay in Ref.~\cite{LHCb-PAPER-2015-034}.
These measurements are used to weight the simulated events used to determine $\varepsilon$ and a systematic uncertainty is assigned as the difference of $\varepsilon$ with and without the weighting.
The full angular distribution of \decay{\Bzb}{\Kstarzb\mumu} has been studied by the LHCb collaboration in Ref.~\cite{LHCb-PAPER-2016-012}.
The decay structure of the \decay{\Bs}{\Kstarzb\mumu} decay is, however, unknown.
To determine a systematic uncertainty associated to the knowledge of these decay models, the simulated samples are weighted such that the coupling strengths used in the model are consistent with the results from global fits to $b \to s$ data~\cite{Altmannshofer:2017fio,Ciuchini:2017mik,Chobanova:2017ghn,Geng:2017svp,Capdevila:2017bsm}.
Again, the systematic uncertainty is assigned as the difference of $\varepsilon$ with and without the weighting.
The total systematic uncertainty due to the knowledge of decay models is 4\% for all measurements.
Finally, the contribution from non-\Kstarzb states in the \decay{\Bs}{\Kstarzb\mumu} is also considered.
This contribution is also unknown and is assumed to be at a similar level as seen in the decay \decay{\Bzb}{\Km\pip\mumu}~\cite{LHCb-PAPER-2016-012}.
Assigning the full size of the effect as systematic uncertainty results in a 3.4\% uncertainty.

The  efficiency ratios used to determine the different branching fraction measurements have an uncertainty of around 1.5\%.
These uncertainties comprise a statistical component due to the limited size of the simulated samples and a systematic component associated to the choice of binning in kinematic variables used to evaluate PID and track reconstruction efficiencies.
A separate systematic uncertainty is also considered on the ratio of efficiencies due to data-simulation differences.
This systematic uncertainty is evaluated by taking the deviation between the efficiency ratio with and without corrections described in Sec.~\ref{sec:Detector} applied.
This includes corrections to the $\B_{(s)}^0$ meson kinematics, PID performance and track reconstruction efficiency.
This results in an additional uncertainty of 1 to 2\% depending on the measurement considered.

%% file: summary.tex
\section{Summary}
\label{sec:Summary}

A search for the decay \decay{\Bs}{\Kstarzb\mumu} is performed using data sets corresponding to 1.0, 2.0 and 1.6\invfb of integrated luminosity collected with the LHCb experiment at centre-of-mass energies of 7, 8 and 13\tev, respectively.
A yield of $38\pm 12$ \decay{\Bs}{\Kstarzb\mumu} decays is obtained, providing the first evidence for this decay with a significance of 3.4 standard deviations above the background-only hypothesis.
The resulting branching fraction is determined to be
\begin{displaymath}
\BF(\decay{\Bs}{\Kstarzb\mumu}) =
[2.9 \pm 1.0\,({\rm stat}) \pm  0.2\,({\rm syst}) \pm 0.3\,({\rm norm})]\times 10^{-8}~.
\end{displaymath}
This measurement is consistent with existing SM predictions of the branching fraction of the decay and a SM-like value of $|V_{td}/V_{ts}|$.
A detailed analysis of the \qsq spectrum of the \decay{\Bs}{\Kstarzb\mumu} decay requires a larger data set.
Such a data set should be available with the upgraded LHCb experiment~\cite{LHCb-TDR-012}.

%% file: acknowledgements.tex
\section*{Acknowledgements}
%
%
\noindent We express our gratitude to our colleagues in the CERN
accelerator departments for the excellent performance of the LHC. We
thank the technical and administrative staff at the LHCb
institutes. We acknowledge support from CERN and from the national
agencies: CAPES, CNPq, FAPERJ and FINEP (Brazil); MOST and NSFC
(China); CNRS/IN2P3 (France); BMBF, DFG and MPG (Germany); INFN
(Italy); NWO (The Netherlands); MNiSW and NCN (Poland); MEN/IFA
(Romania); MinES and FASO (Russia); MinECo (Spain); SNSF and SER
(Switzerland); NASU (Ukraine); STFC (United Kingdom); NSF (USA).  We
acknowledge the computing resources that are provided by CERN, IN2P3
(France), KIT and DESY (Germany), INFN (Italy), SURF (The
Netherlands), PIC (Spain), GridPP (United Kingdom), RRCKI and Yandex
LLC (Russia), CSCS (Switzerland), IFIN-HH (Romania), CBPF (Brazil),
PL-GRID (Poland) and OSC (USA). We are indebted to the communities
behind the multiple open-source software packages on which we depend.
Individual groups or members have received support from AvH Foundation
(Germany), EPLANET, Marie Sk\l{}odowska-Curie Actions and ERC
(European Union), ANR, Labex P2IO and OCEVU, and R\'{e}gion
Auvergne-Rh\^{o}ne-Alpes (France), Key Research Program of Frontier
Sciences of CAS, CAS PIFI, and the Thousand Talents Program (China),
RFBR, RSF and Yandex LLC (Russia), GVA, XuntaGal and GENCAT (Spain),
Herchel Smith Fund, the Royal Society, the English-Speaking Union and
the Leverhulme Trust (United Kingdom).

%% file: appendix.tex
\clearpage


\appendix

\section*{Appendix}

\begin{figure}[!htb]
\begin{minipage}[c]{0.70\textwidth}
\noindent
{\small
\fcolorbox{black}{colour:sig}{\phantom{-}}~$\Bs\to\jpsi\Kstarzb$~
\fcolorbox{black}{colour:Kstmm}{\phantom{-}}~$\Bzb \to \jpsi\Kstarzb$~
\fcolorbox{black}{colour:pKmm}{\phantom{-}}~$\Lb \to \jpsi p \Km$ \\
\fcolorbox{black}{colour:Kmm}{\phantom{-}}~$\Bp \to \jpsi \Kp$~
\fcolorbox{black}{colour:combinatorial}{\phantom{-}}~combinatorial background~
$\bf{-}$~fit~
$\bullet$~data
}
\end{minipage}
\vspace{2mm}
\centering
\includegraphics[width=0.4\linewidth]{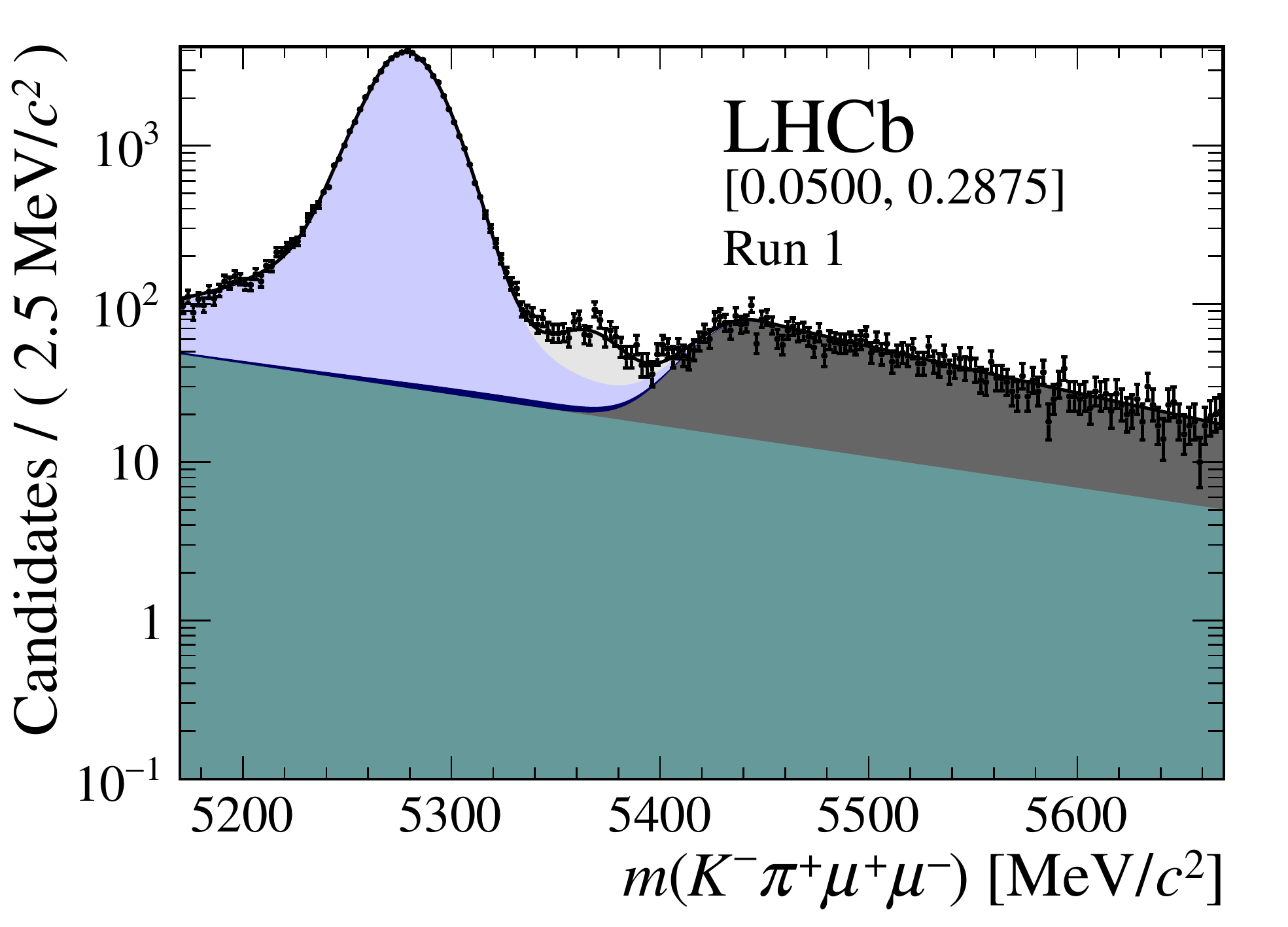}
\includegraphics[width=0.4\linewidth]{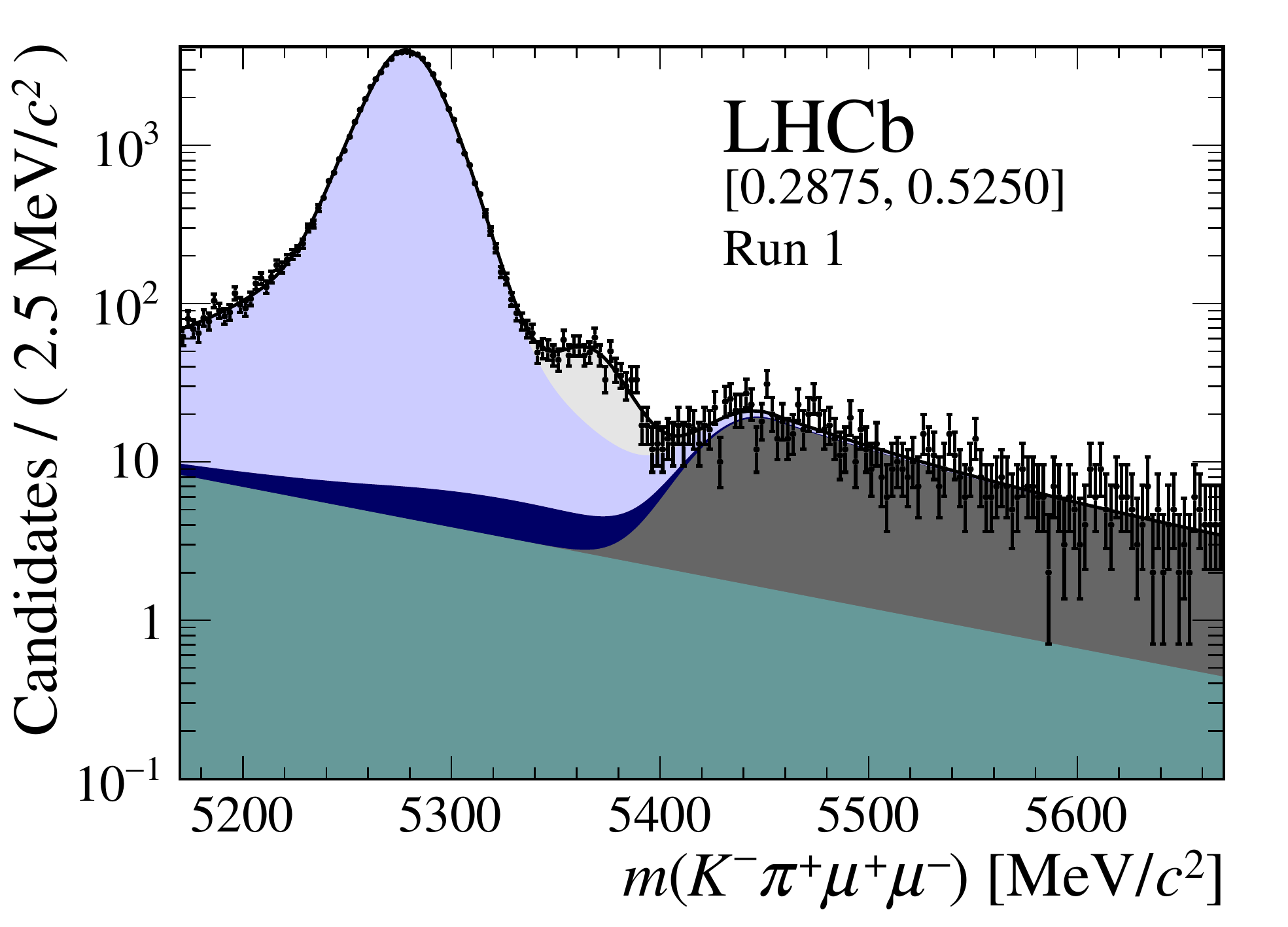} \\
\includegraphics[width=0.4\linewidth]{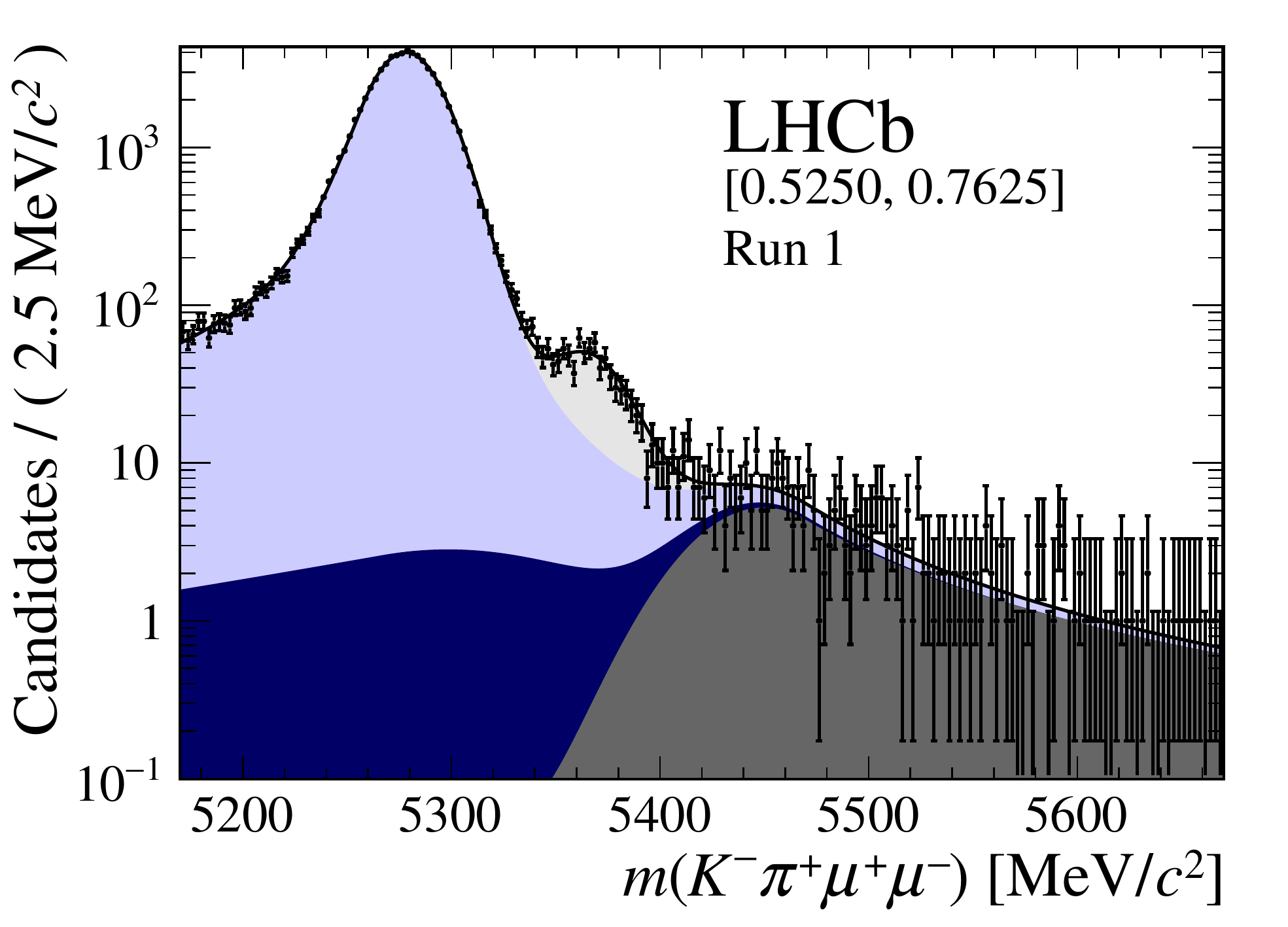}
\includegraphics[width=0.4\linewidth]{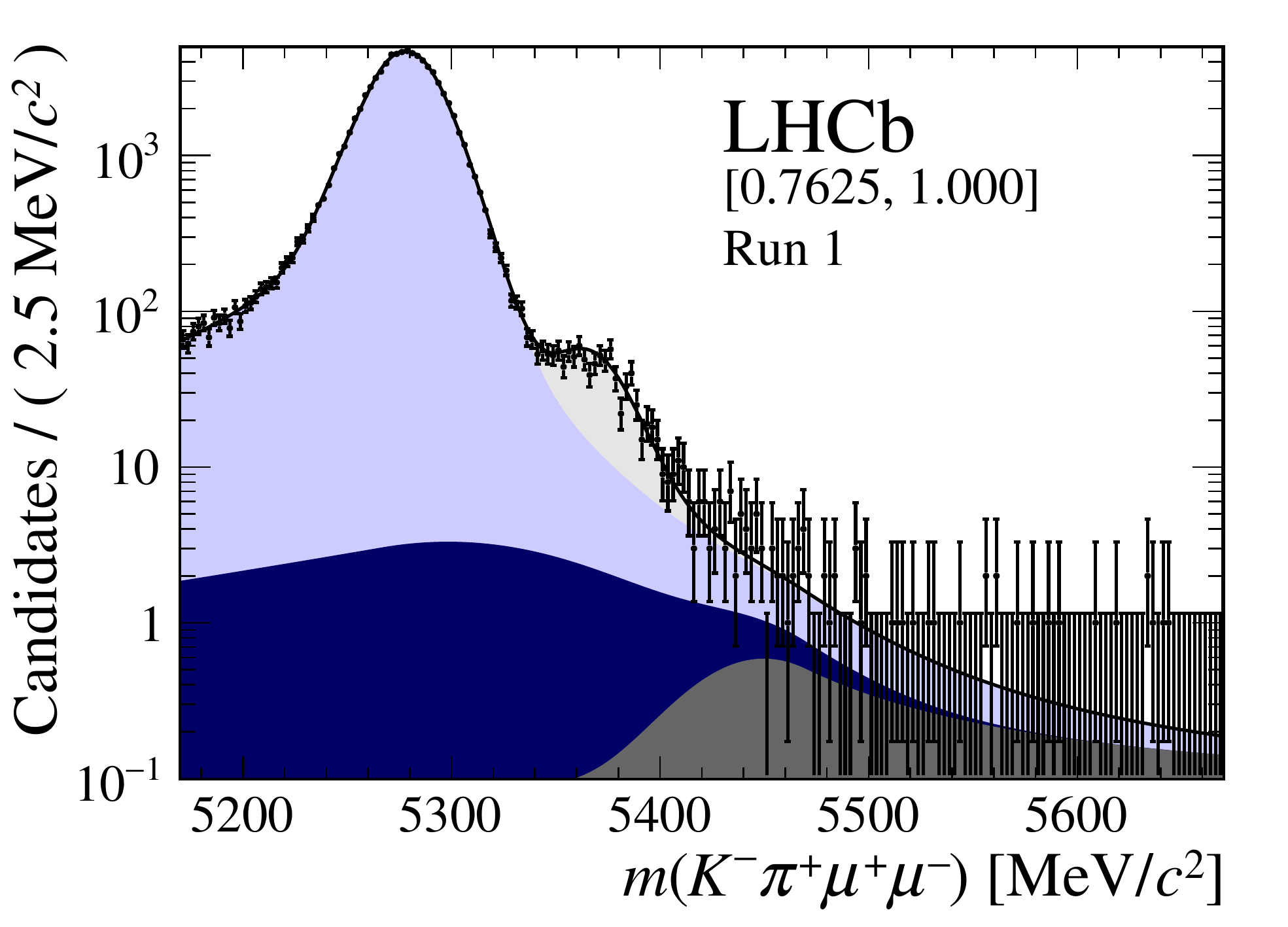} \\
\includegraphics[width=0.4\linewidth]{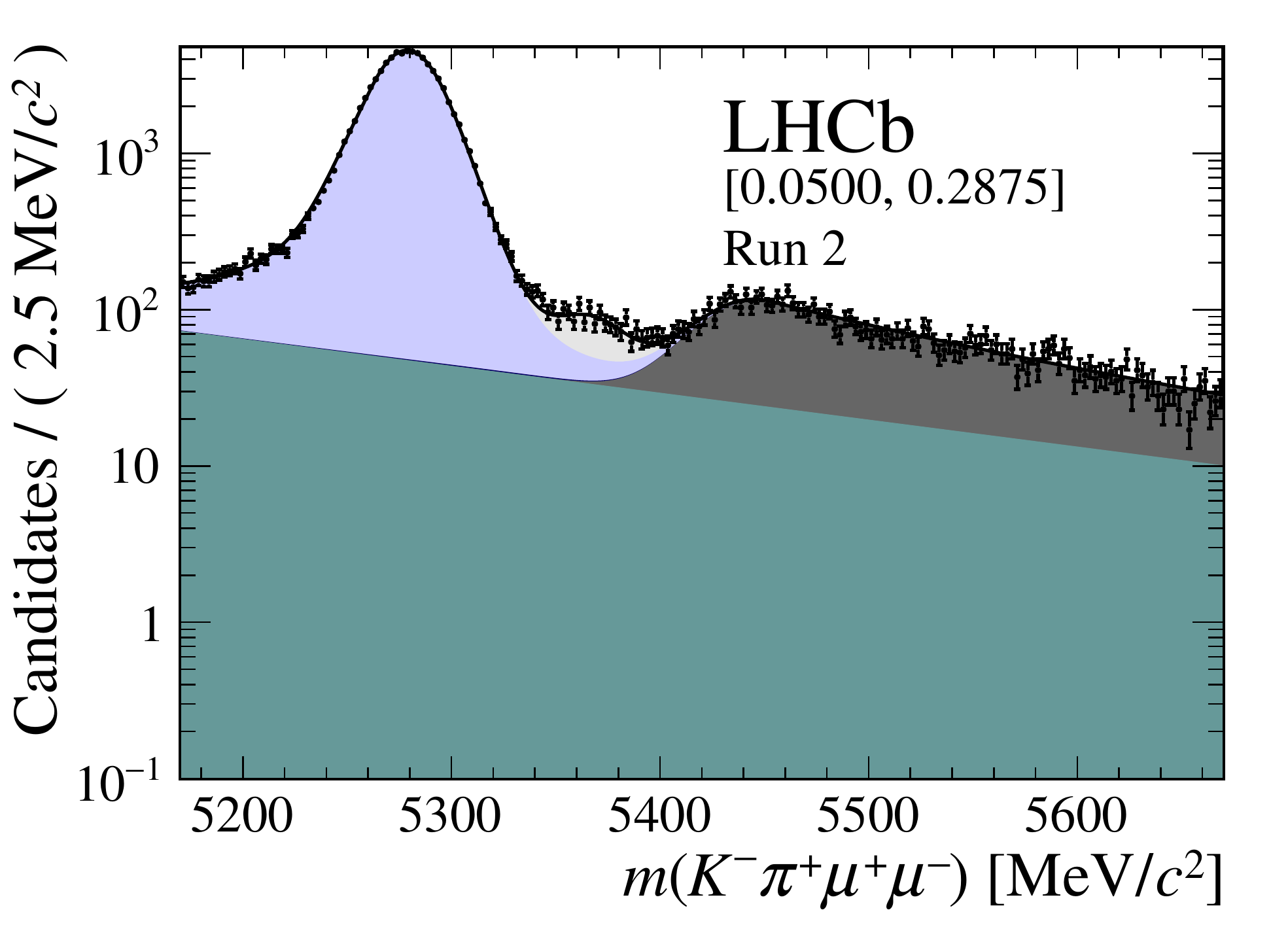}
\includegraphics[width=0.4\linewidth]{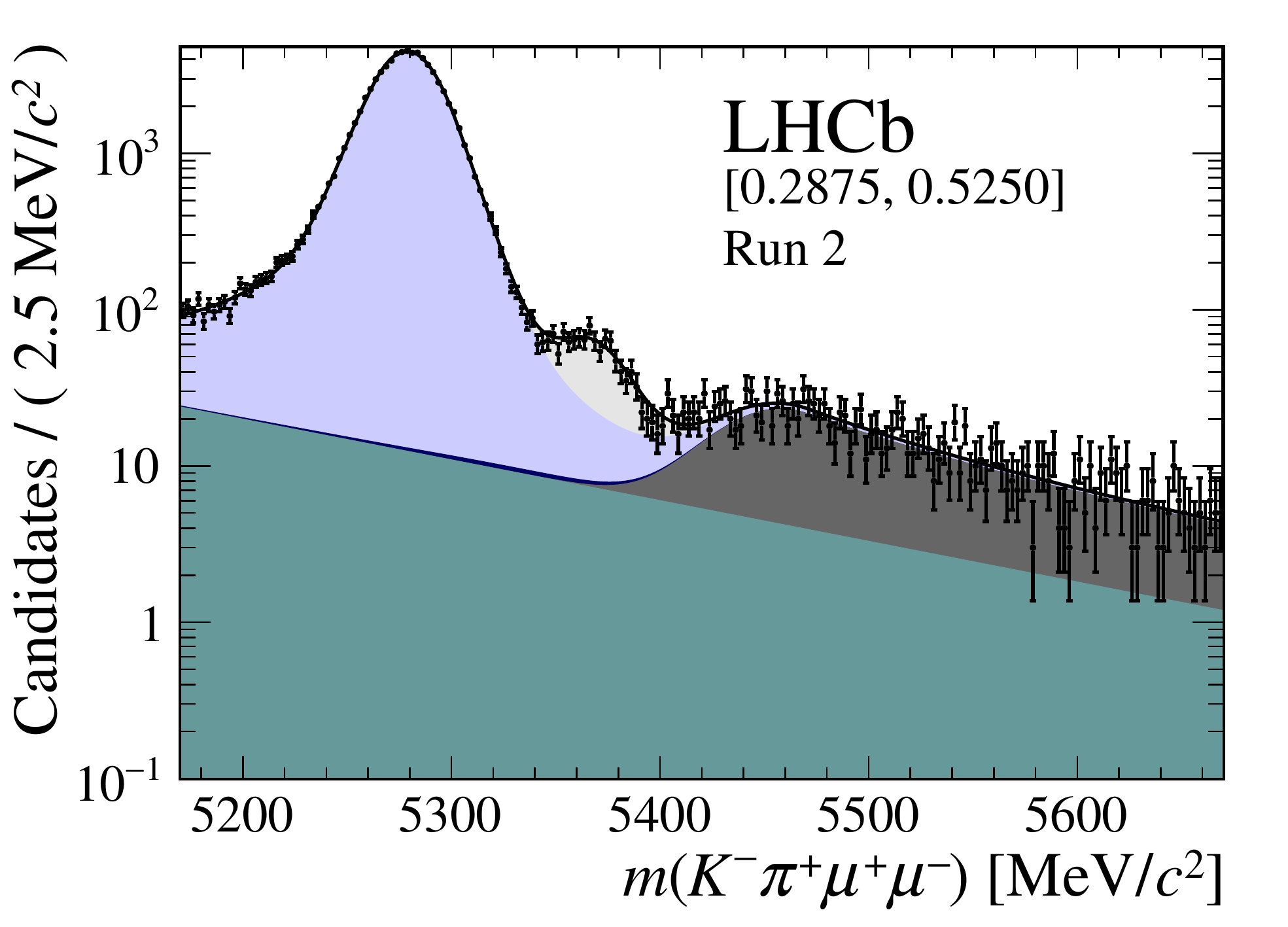} \\
\includegraphics[width=0.4\linewidth]{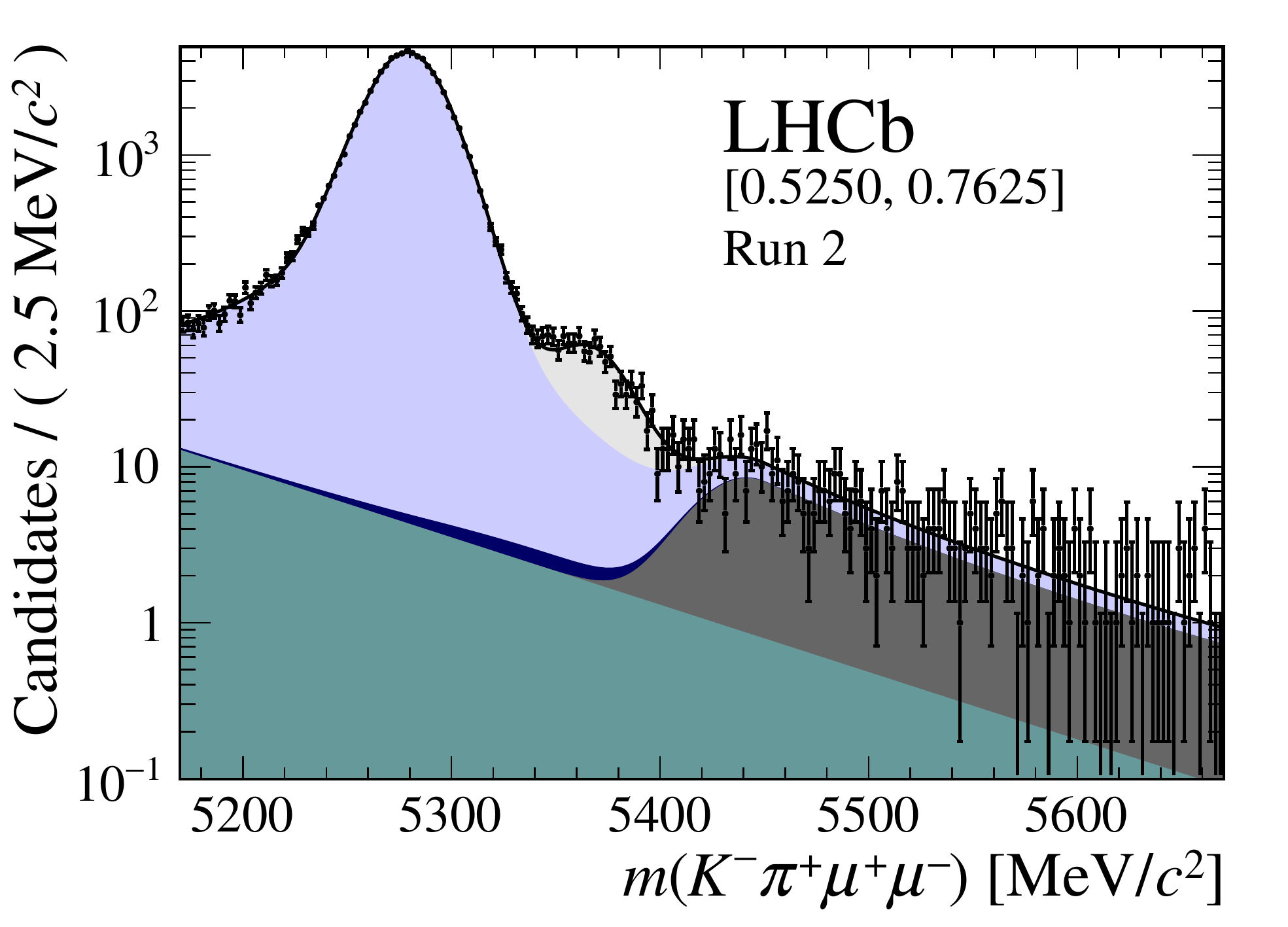}
\includegraphics[width=0.4\linewidth]{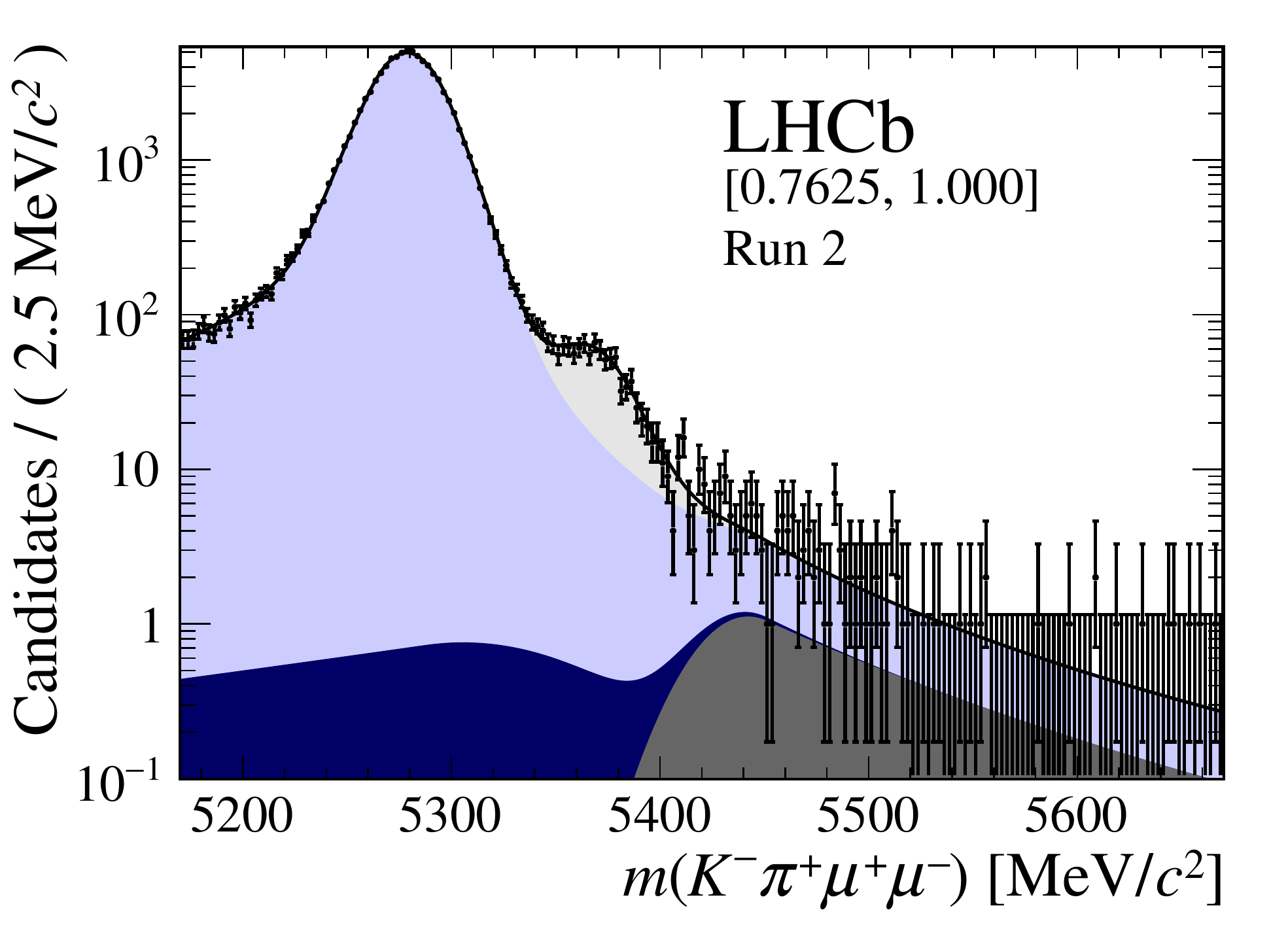} \\
\caption{
Distribution of reconstructed $\Km\pip\mumu$ invariant mass of candidates in the \jpsi mass window in (top four figures) the Run~1 and (bottom four figures) Run~2  data sets.
The candidates are divided into four independent bins of increasing neural network response per data taking period.
}
\label{fig:appendix:fits:jpsinocon}
\end{figure}

\begin{figure}[!htb]
\begin{minipage}[c]{0.70\textwidth}
\noindent
{\small
\fcolorbox{black}{colour:sig}{\phantom{-}}~$\Bs\to\Kstarzb\mumu$~
\fcolorbox{black}{colour:Kstmm}{\phantom{-}}~$\Bzb \to \Kstarzb\mumu$~
\fcolorbox{black}{colour:pKmm}{\phantom{-}}~$\Lb \to p \Km \mumu$ \\
\fcolorbox{black}{colour:combinatorial}{\phantom{-}}~combinatorial background~
$\bf{-}$~fit~
$\bullet$~data
}
\end{minipage}
\vspace{2mm}
\centering
\includegraphics[width=0.4\linewidth]{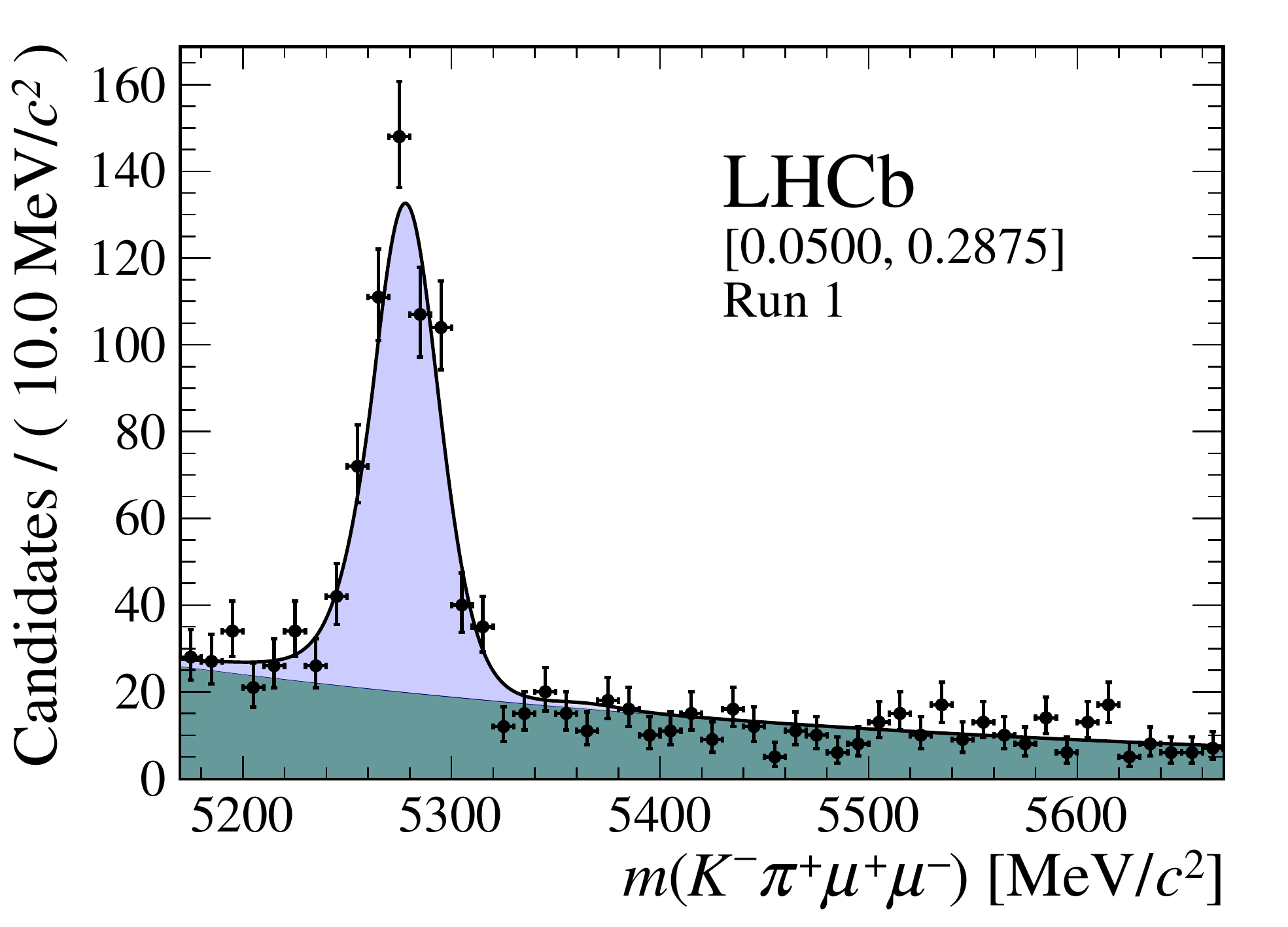}
\includegraphics[width=0.4\linewidth]{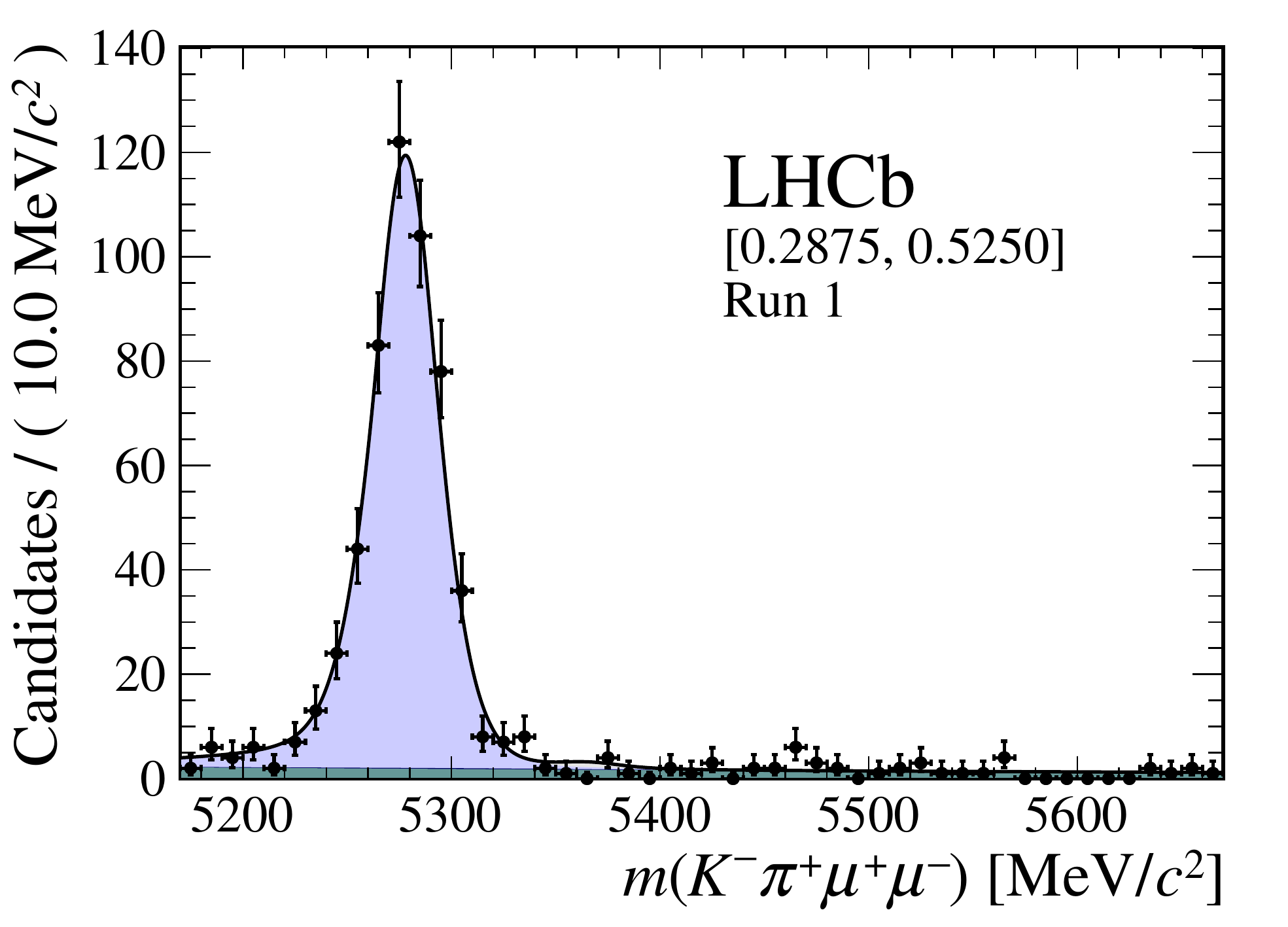} \\
\includegraphics[width=0.4\linewidth]{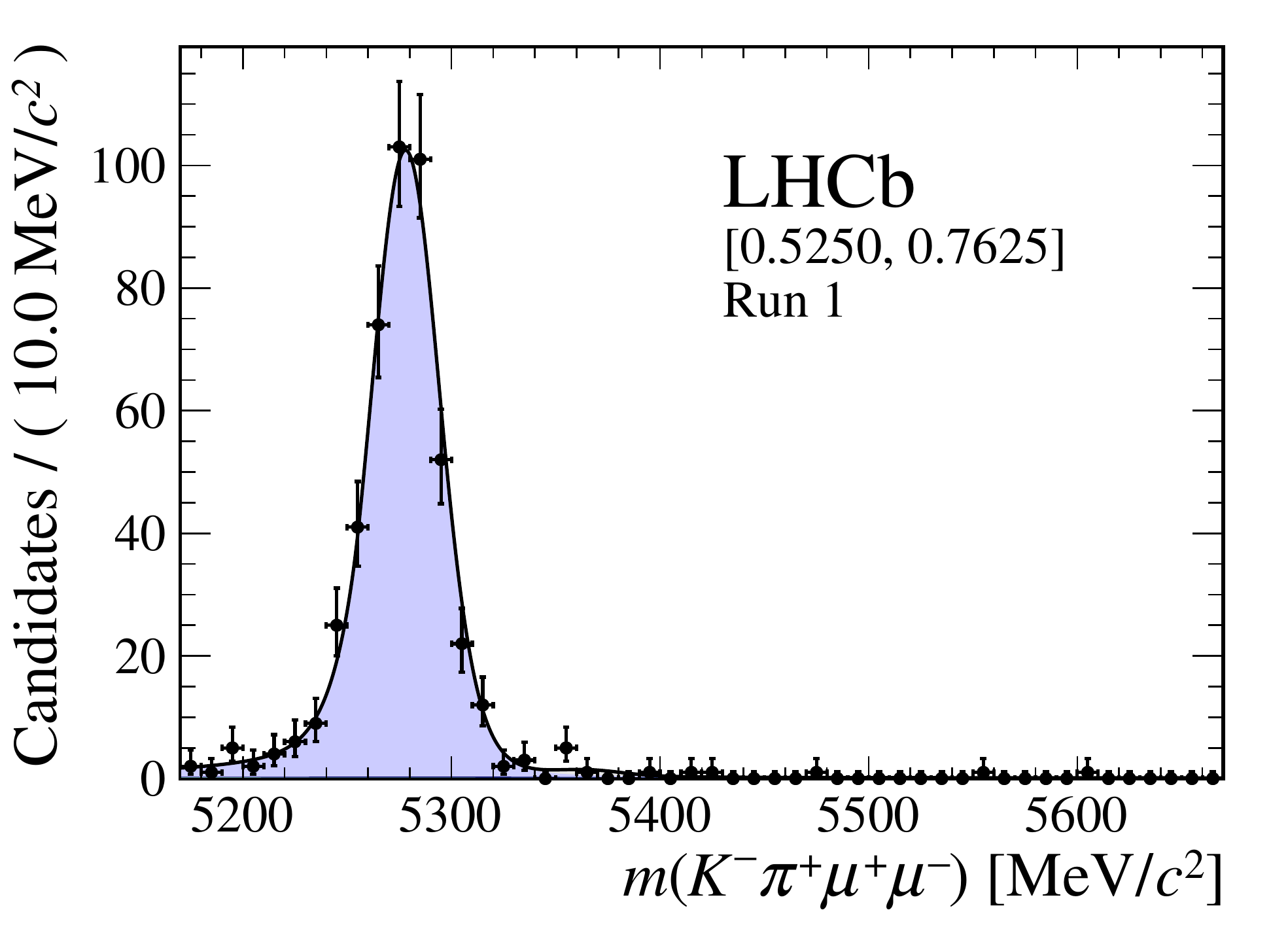}
\includegraphics[width=0.4\linewidth]{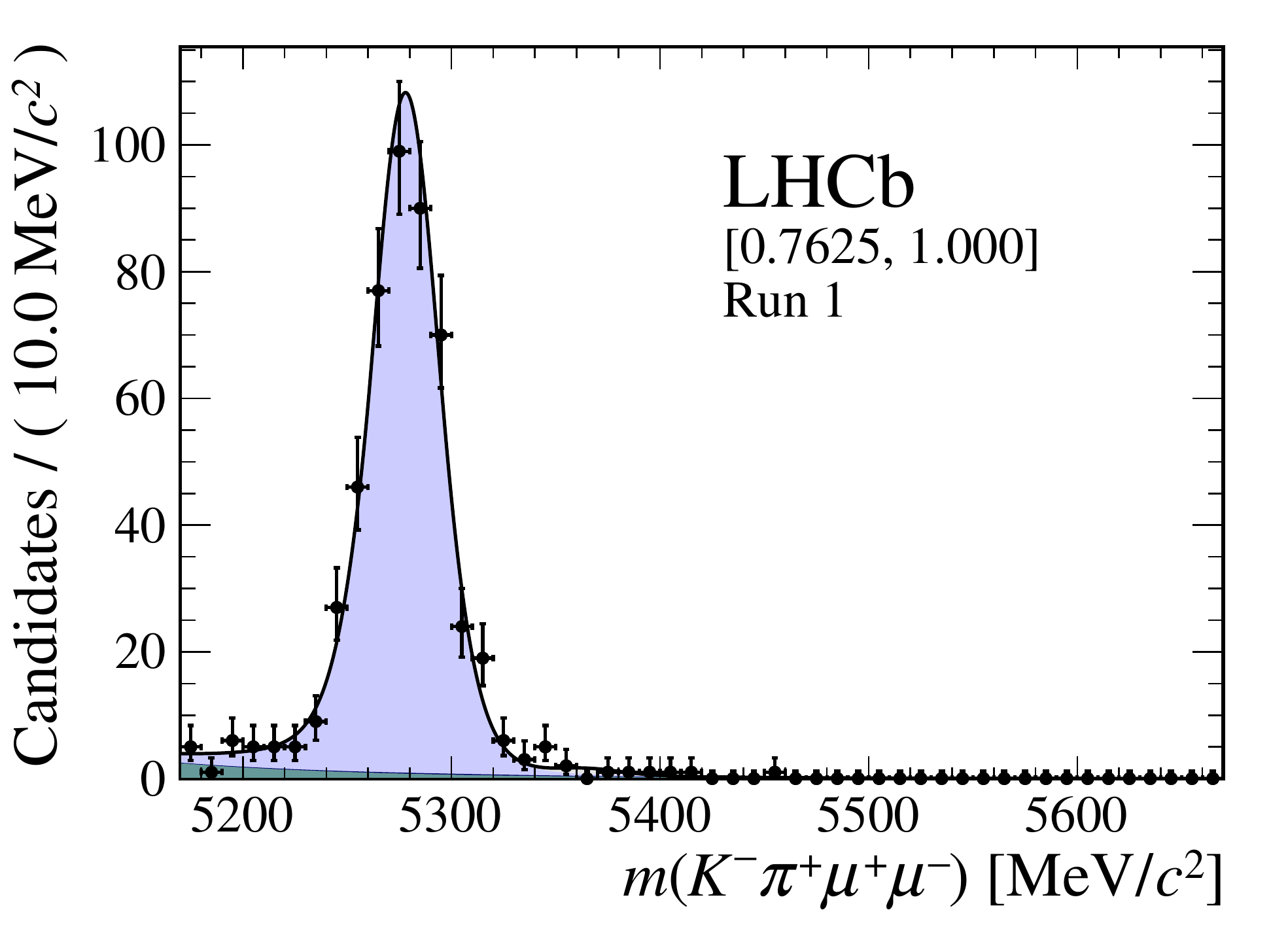} \\
\includegraphics[width=0.4\linewidth]{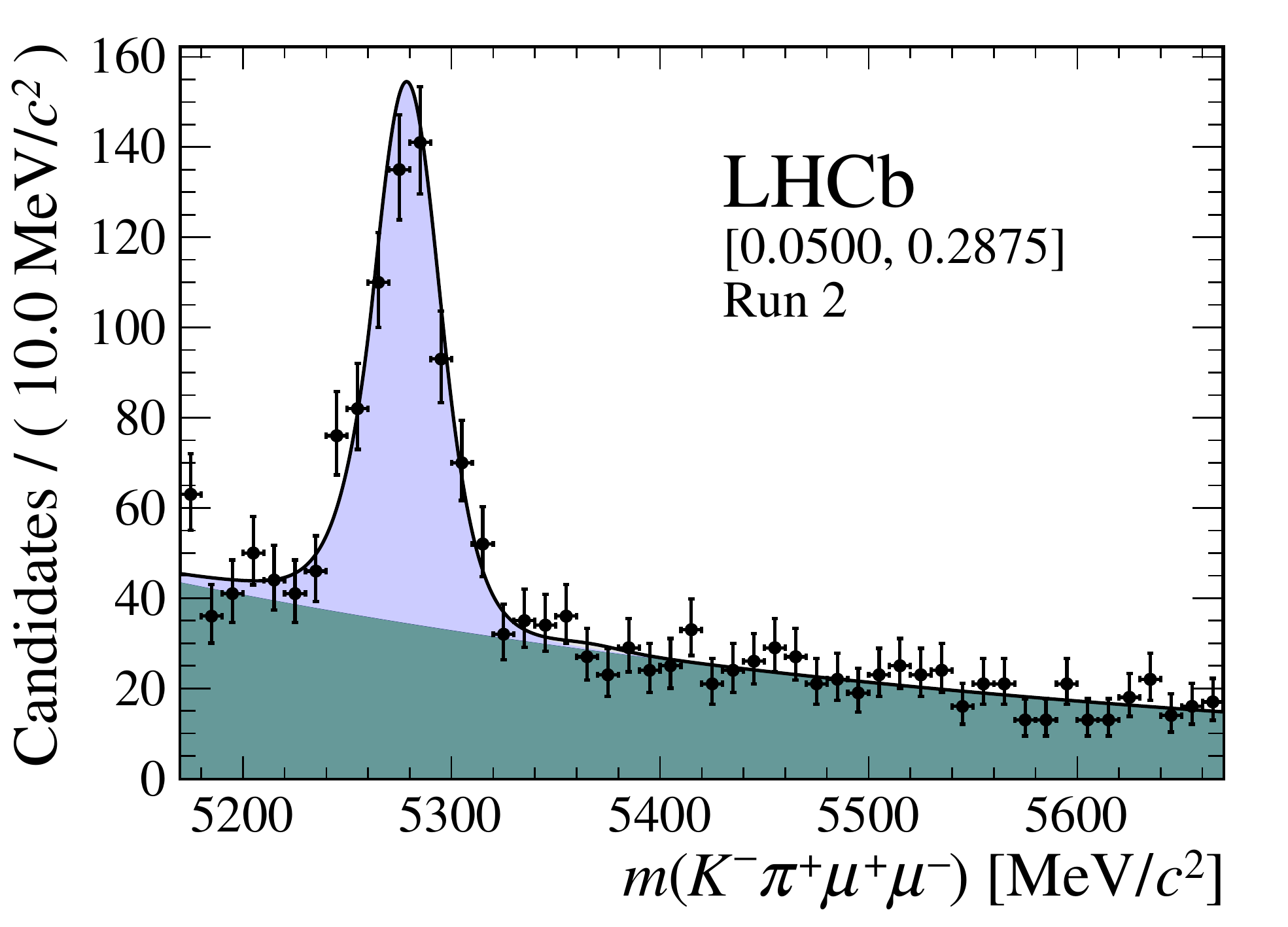}
\includegraphics[width=0.4\linewidth]{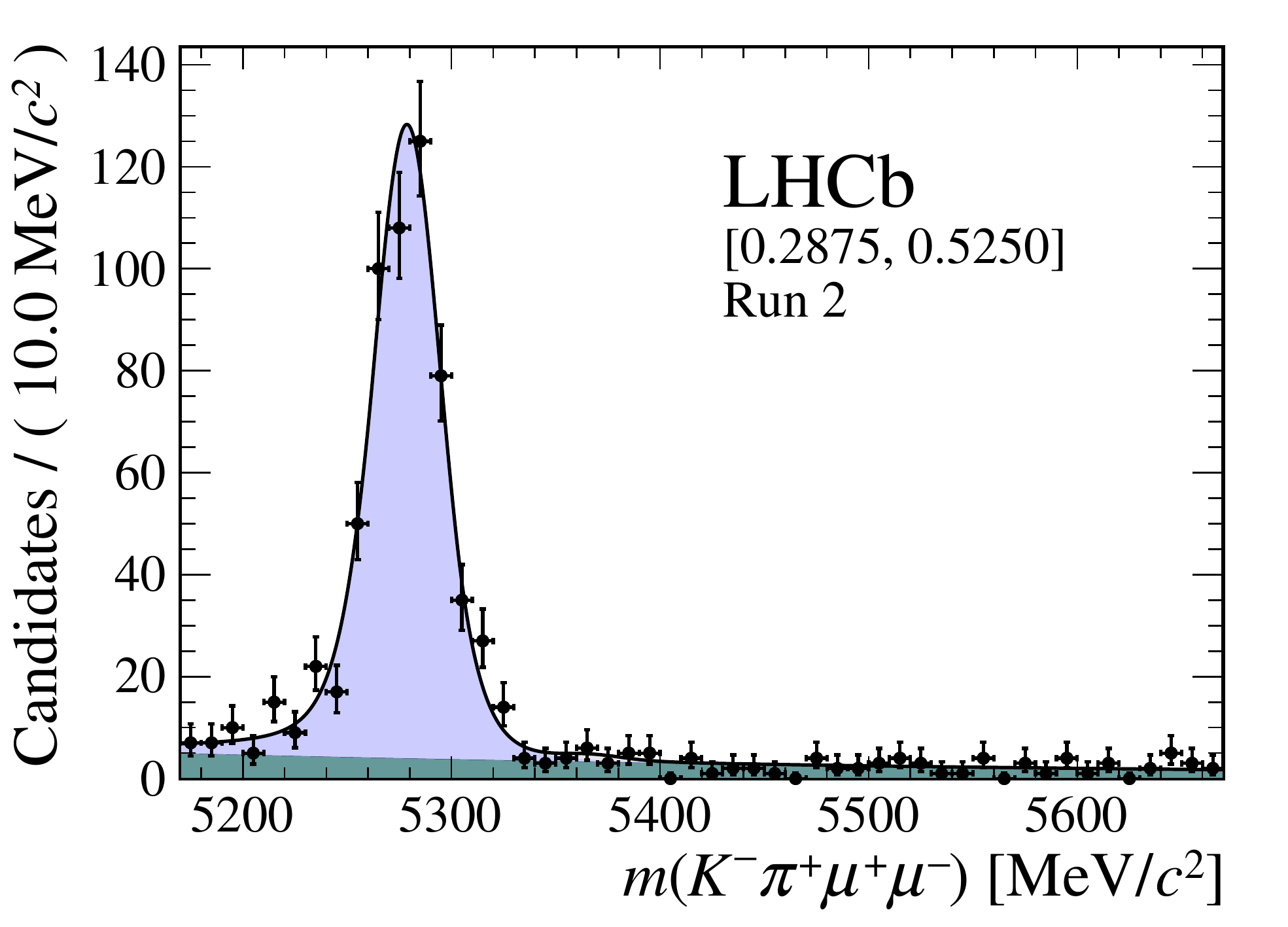} \\
\includegraphics[width=0.4\linewidth]{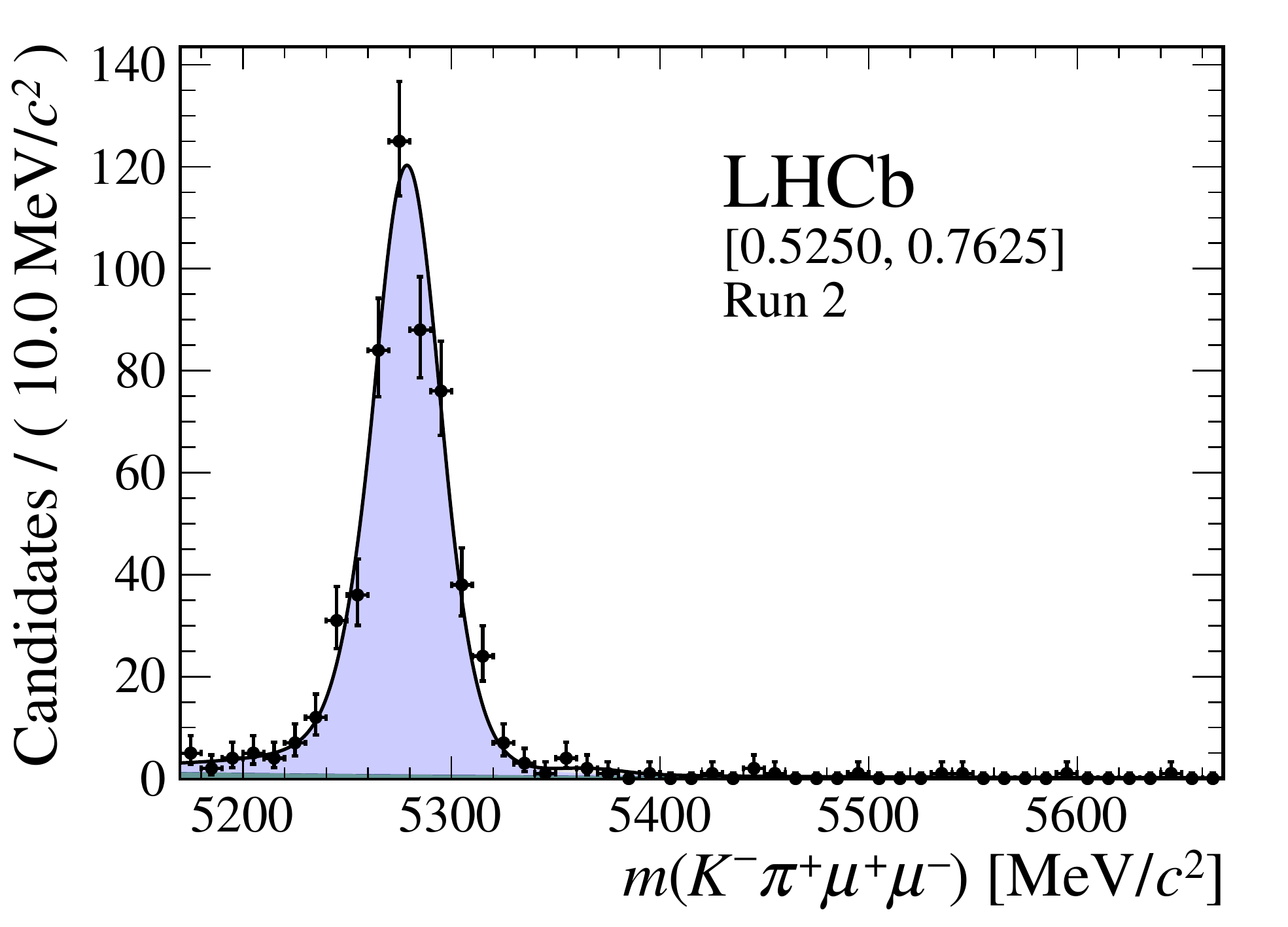}
\includegraphics[width=0.4\linewidth]{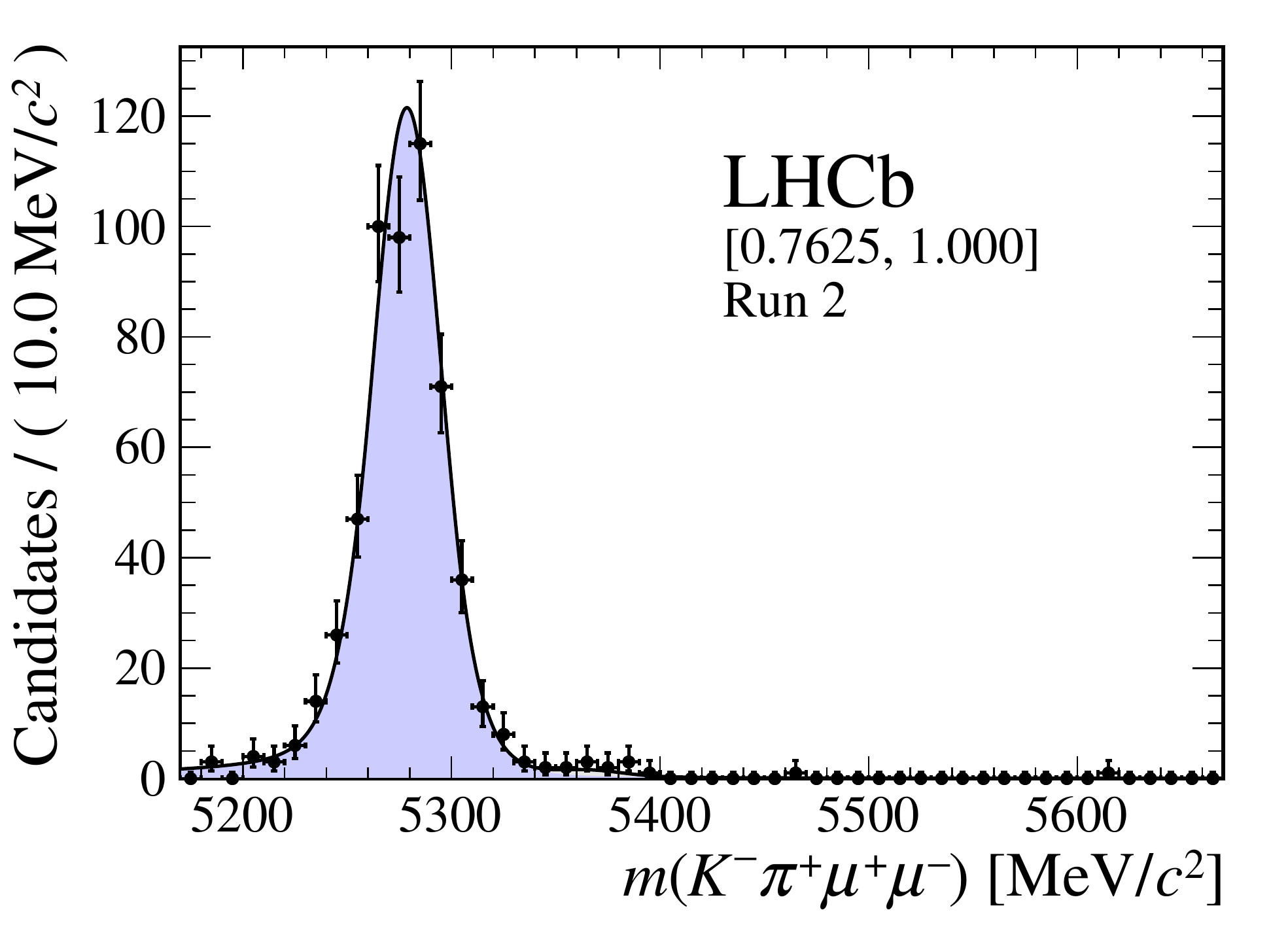} \\
\caption{
Distribution of reconstructed $\Km\pip\mumu$ invariant mass of candidates outside of the \jpsi and \psitwos mass regions in (top four figures)  the Run~1 and  (bottom four figures) Run~2 data sets.
The candidates are divided into four independent bins of increasing neural network response per data taking period.
}
\label{fig:appendix:fits:mumu}
\end{figure}

\begin{figure}[!htb]
\begin{minipage}[c]{0.80\textwidth}
\noindent
{\small
\fcolorbox{black}{colour:sig}{\phantom{-}}~$\Bs\to\jpsi\Kstarzb$~
\fcolorbox{black}{colour:Kstmm}{\phantom{-}}~$\Bzb \to \jpsi\Kstarzb$~
\fcolorbox{black}{colour:Kstmm:bkg}{\phantom{-}}~$\Bzb \to\Kstarzb\mumu$~
\fcolorbox{black}{colour:pKmm}{\phantom{-}}~$\Lb \to \jpsi p \Km$ \\
\fcolorbox{black}{colour:Kmm}{\phantom{-}}~$\Bp \to \jpsi \Kp$~
\fcolorbox{black}{colour:combinatorial}{\phantom{-}}~combinatorial background~
$\bf{-}$~fit~
$\bullet$~data
}
\vspace{2mm}
\end{minipage}
\centering
\includegraphics[width=0.4\linewidth]{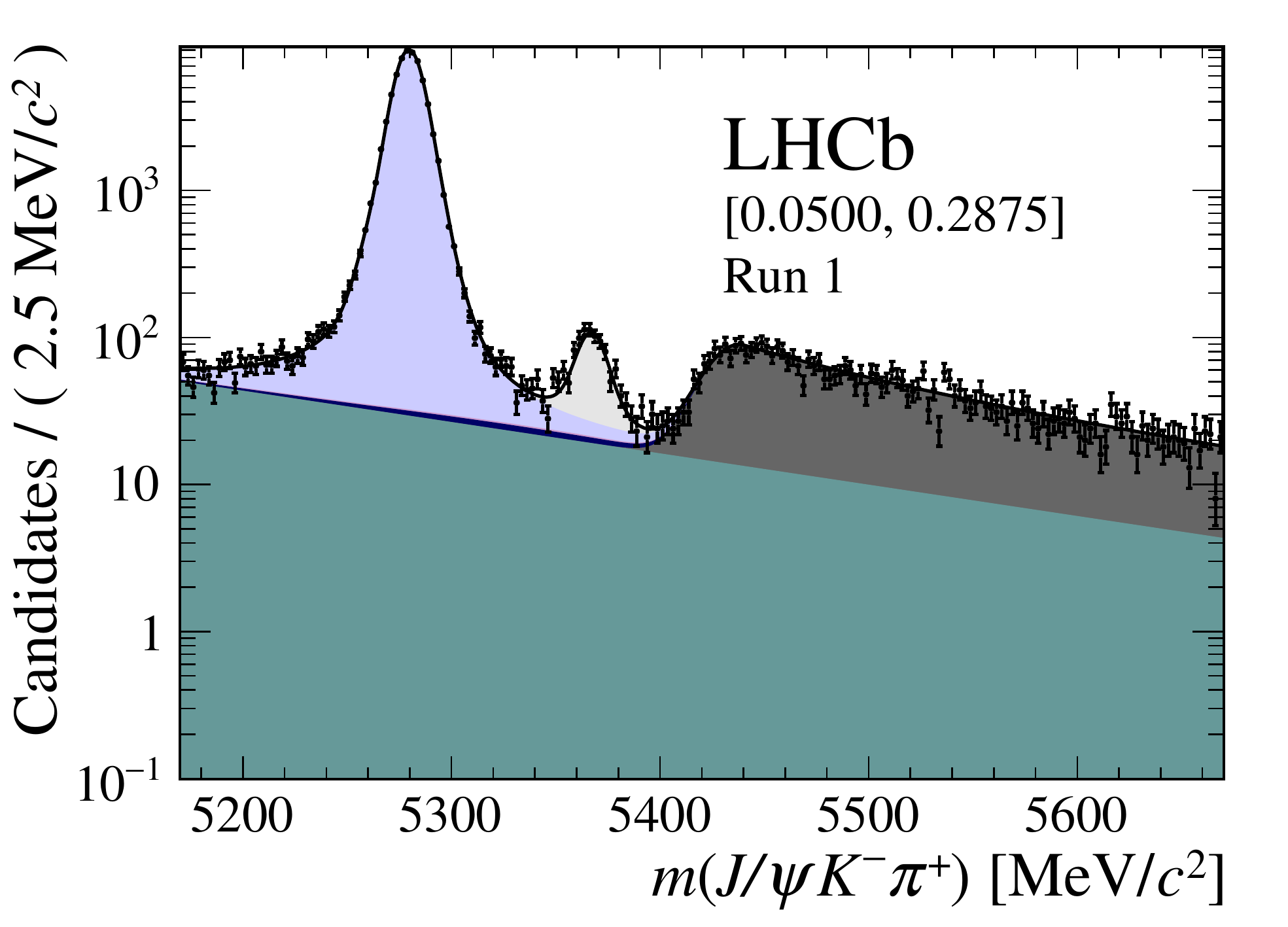}
\includegraphics[width=0.4\linewidth]{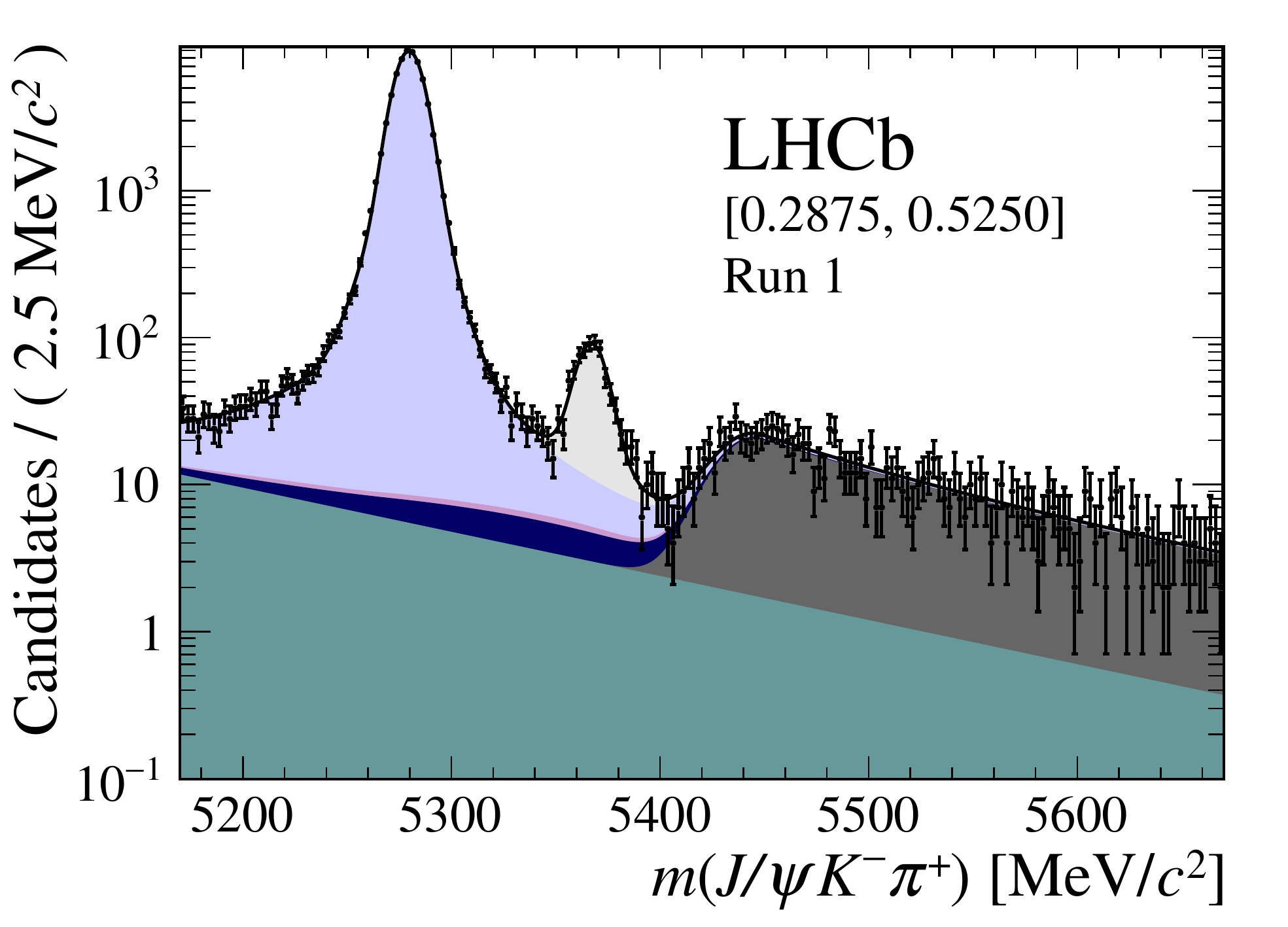}
\includegraphics[width=0.4\linewidth]{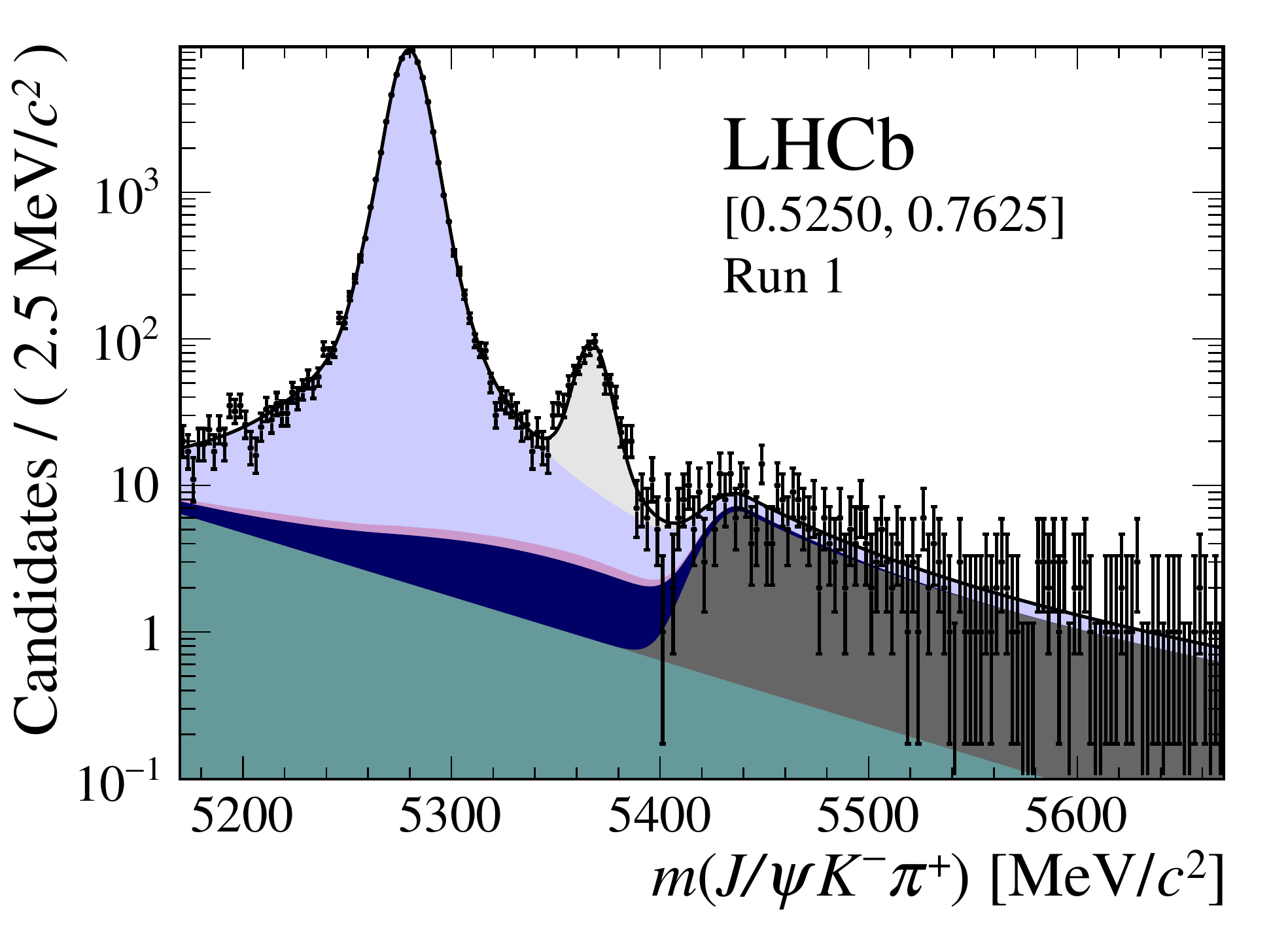}
\includegraphics[width=0.4\linewidth]{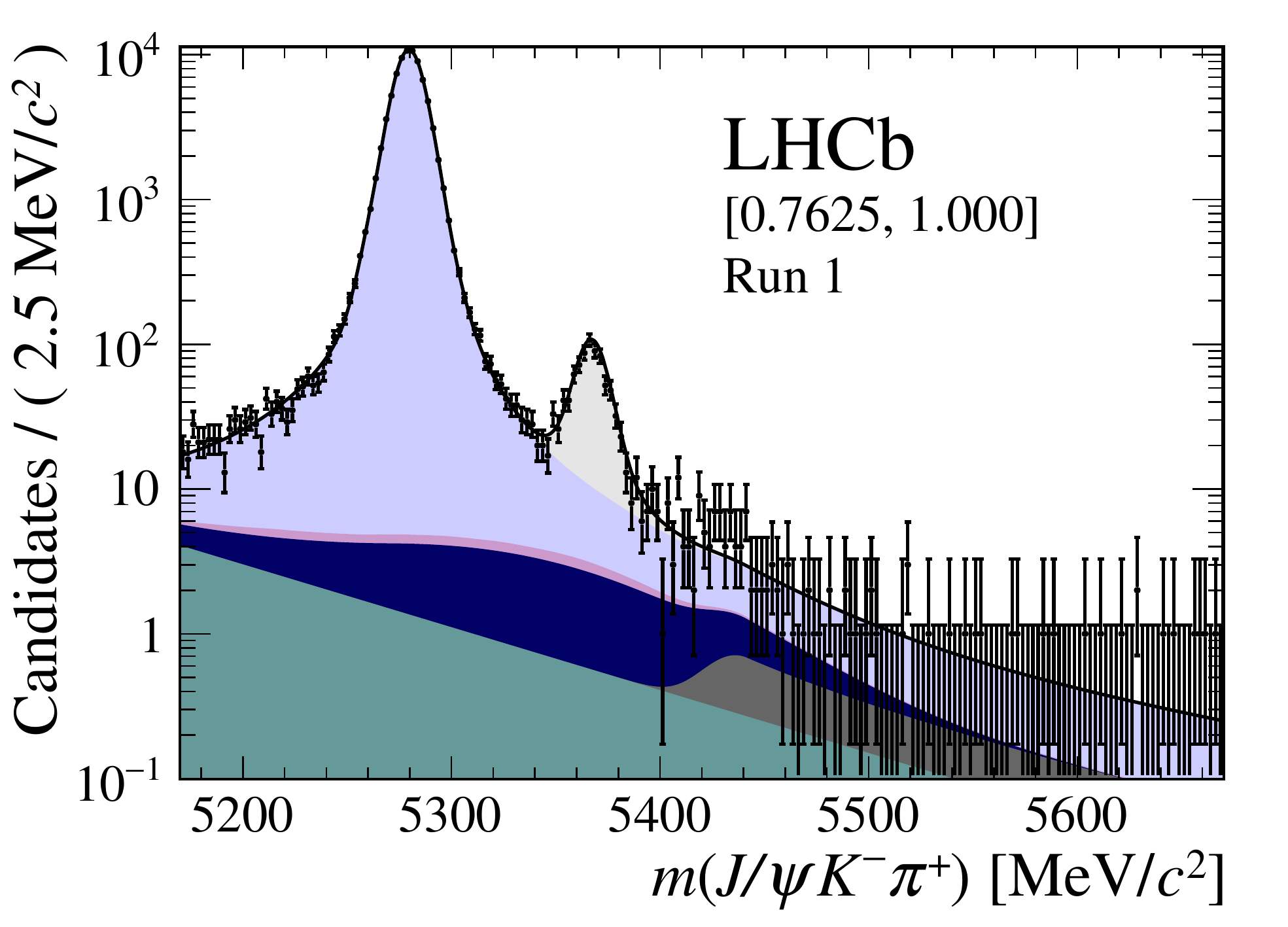} \\
\includegraphics[width=0.4\linewidth]{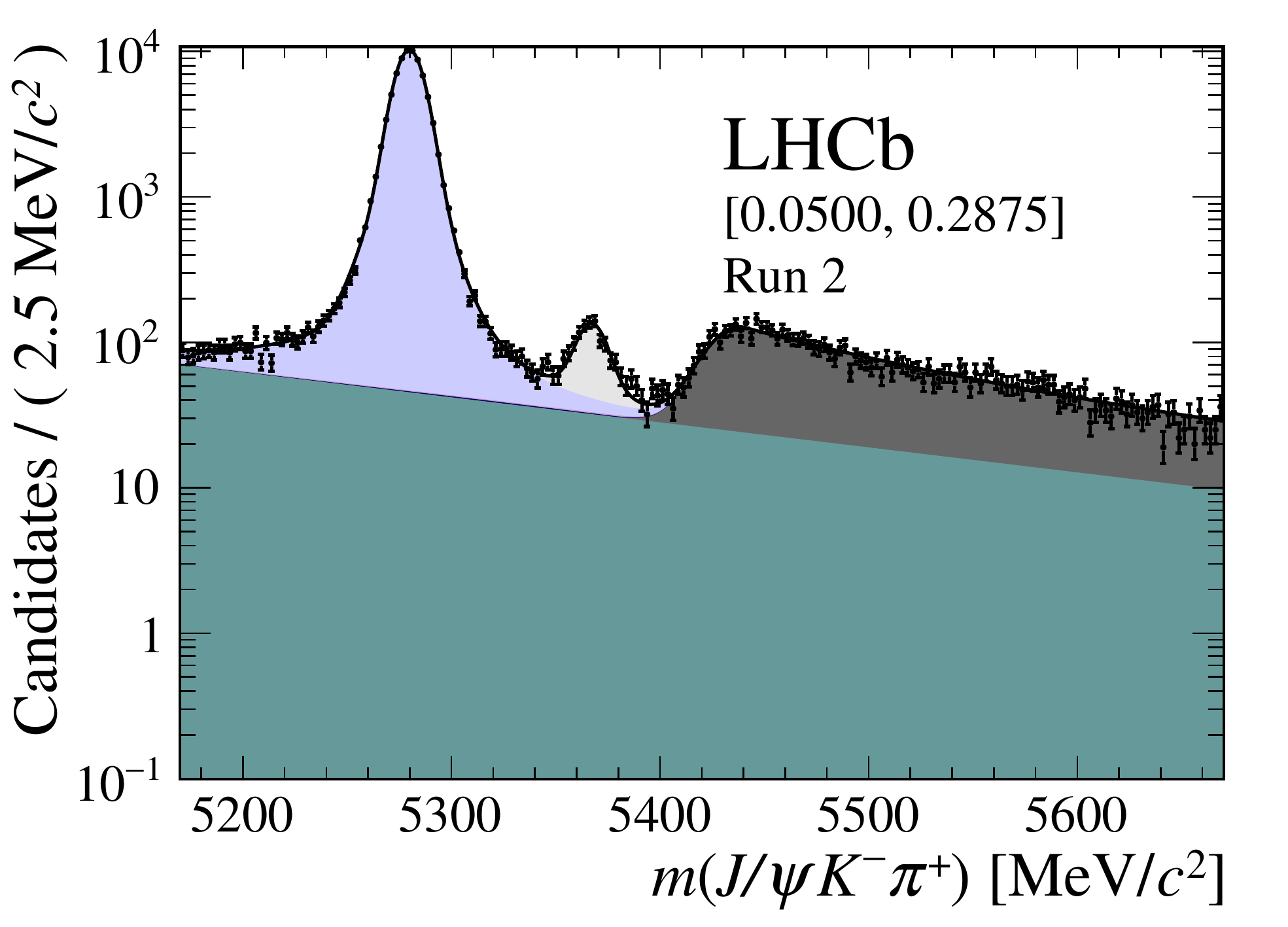}
\includegraphics[width=0.4\linewidth]{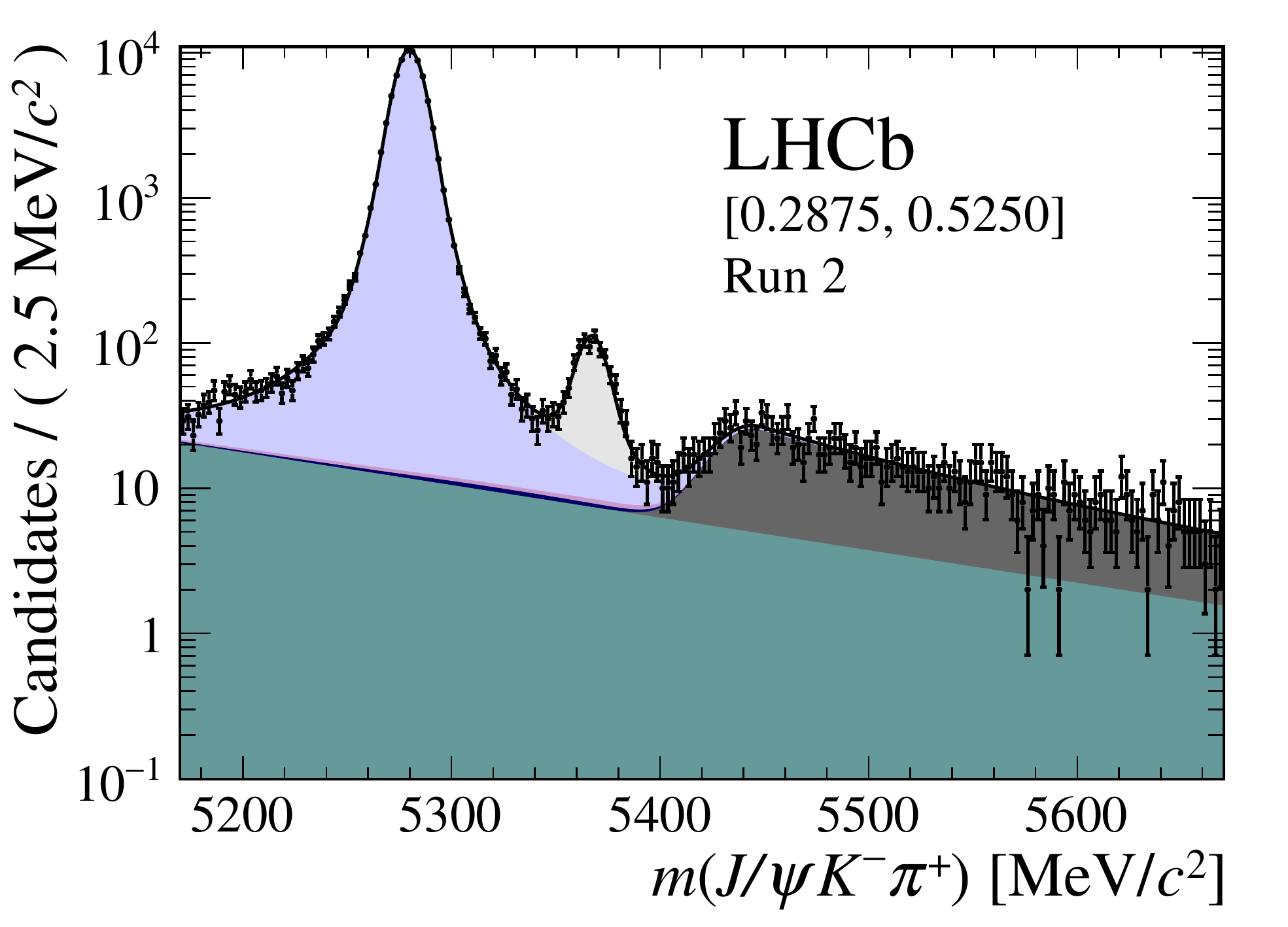}
\includegraphics[width=0.4\linewidth]{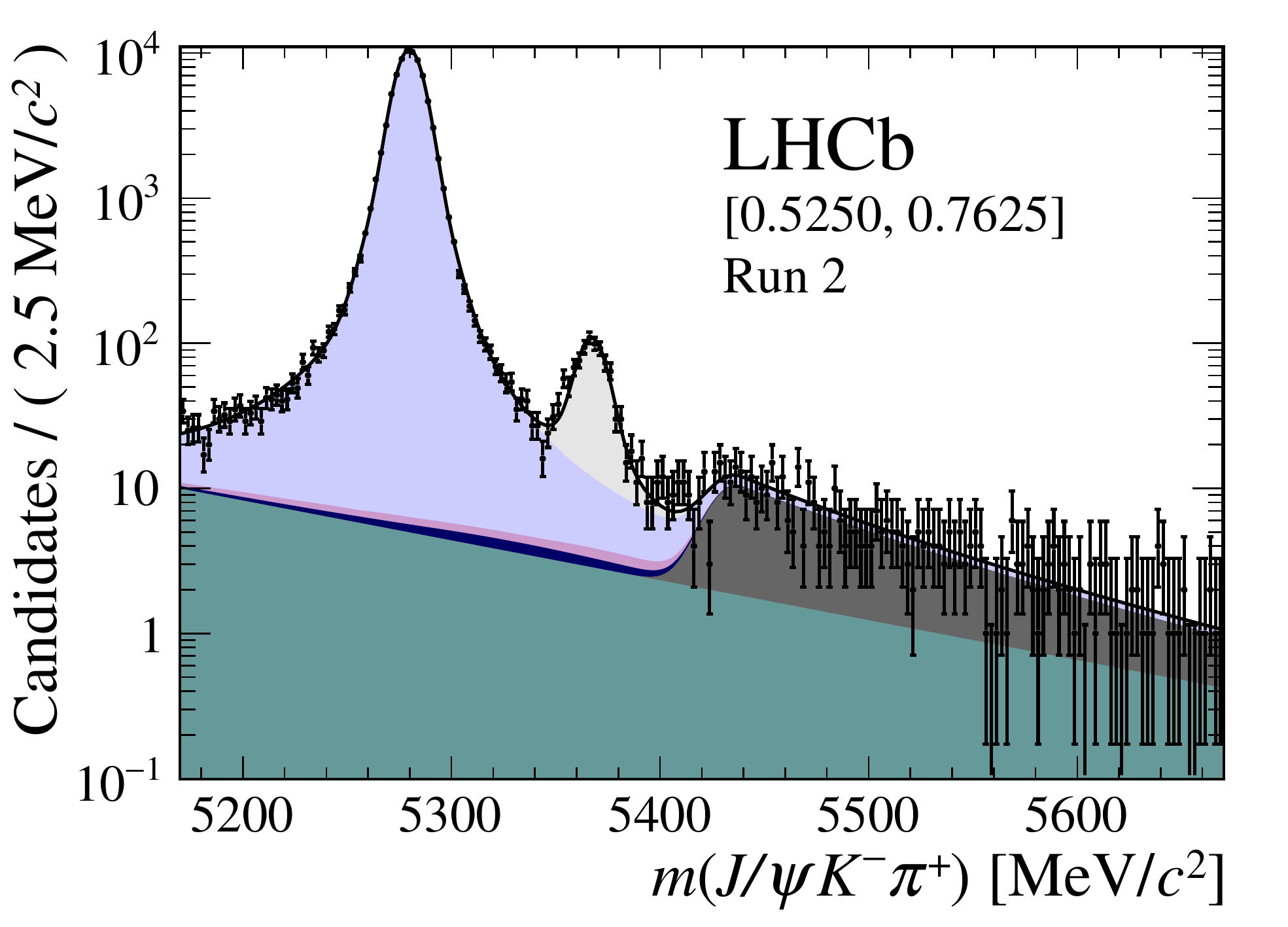}
\includegraphics[width=0.4\linewidth]{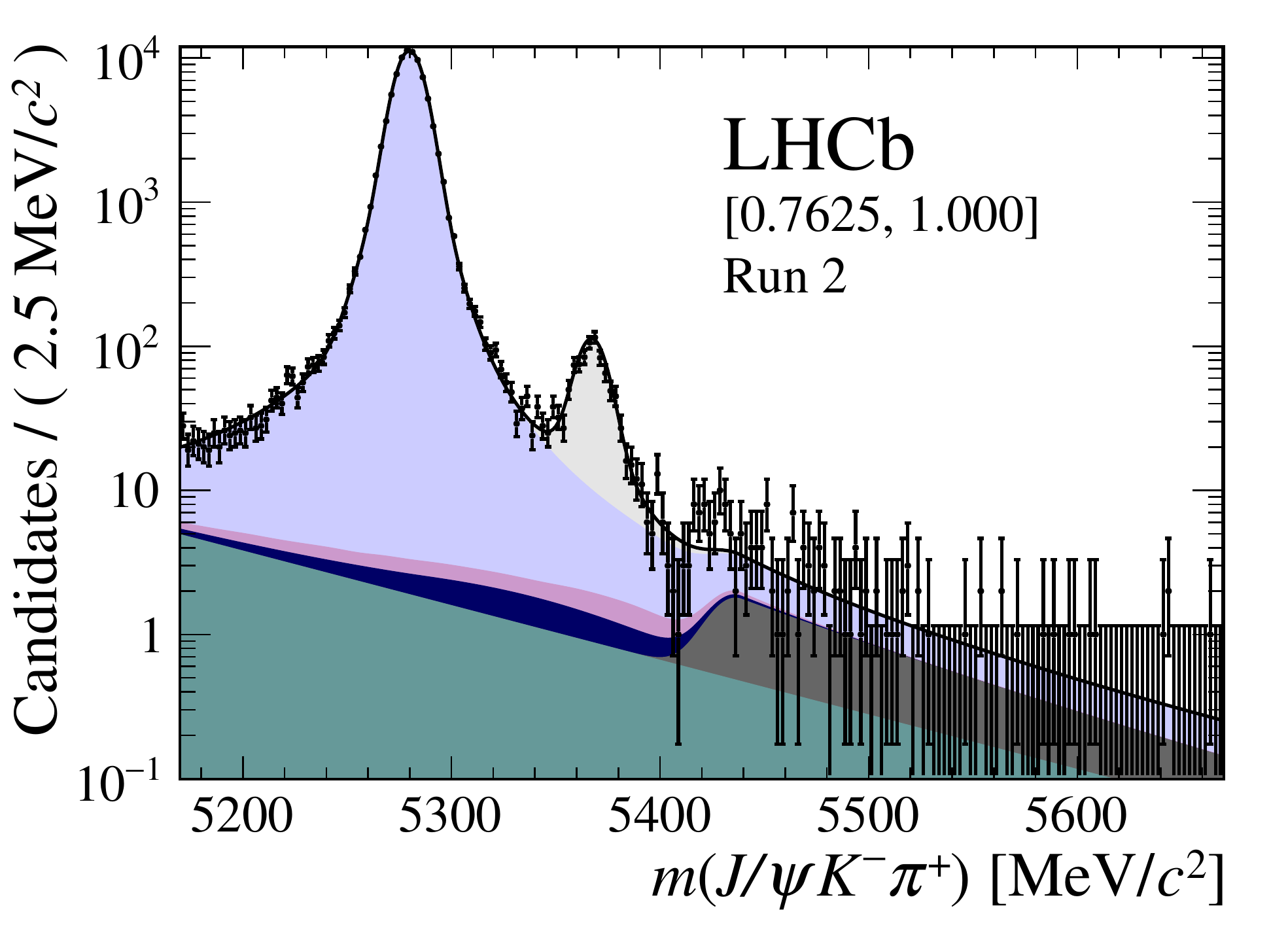} \\
\caption{
Distribution of reconstructed $\jpsi\Km\pip$ invariant mass after application of a \jpsi mass constraint of candidates in (top four figures)  the Run~1 and (bottom four figures)  Run~2 data sets.
The candidates are divided into four independent bins of increasing neural network response per data taking period.
}
\label{fig:appendix:fits:jpsi}
\end{figure}

\noindent
In these appendices, the fits to the $\jpsi\Km\pip$ and $\Km\pip\mumu$ invariant mass of the selected candidates in bins of neural network response for both the Run~1 and Run~2 data sets are shown.
The fit to the $\Km\pip\mumu$ invariant mass of the candidates in the \jpsi mass window is shown in Fig.~\ref{fig:appendix:fits:jpsinocon}.
This fit is used to determine the resolution and tail parameters for the \decay{\Bs}{\Kstarzb\mumu} decay.
The fit to $\Km\pip\mumu$ invariant mass of the \decay{\Bs}{\Kstarzb\mumu} candidates is shown in Fig.~\ref{fig:appendix:fits:mumu}.
The fit to the $\jpsi\Km\pip$ invariant mass after application of the \jpsi mass constraint is shown in Fig.~\ref{fig:appendix:fits:jpsi}.

\clearpage

\clearpage

\clearpage

%% file: LHCb_Authorship_flat_05-Feb-2018.tex
\centerline{\large\bf LHCb collaboration}
\begin{flushleft}
\small
R.~Aaij$^{43}$,
B.~Adeva$^{39}$,
M.~Adinolfi$^{48}$,
Z.~Ajaltouni$^{5}$,
S.~Akar$^{59}$,
P.~Albicocco$^{19}$,
J.~Albrecht$^{10}$,
F.~Alessio$^{40}$,
M.~Alexander$^{53}$,
A.~Alfonso~Albero$^{38}$
% ...

{\footnotesize \it
% ...
$ ^{5}$Clermont Universit{\'e}, Universit{\'e} Blaise Pascal, CNRS/IN2P3, LPC, Clermont-Ferrand, France\\
% ...
$ ^{10}$Fakult{\"a}t Physik, Technische Universit{\"a}t Dortmund, Dortmund, Germany\\
% ...
$ ^{19}$Laboratori Nazionali dell'INFN di Frascati, Frascati, Italy\\
% ...
$ ^{38}$ICCUB, Universitat de Barcelona, Barcelona, Spain\\
$ ^{39}$Instituto Galego de F{\'\i}sica de Altas Enerx{\'\i}as (IGFAE), Universidade de Santiago de Compostela, Santiago de Compostela, Spain\\
$ ^{40}$European Organization for Nuclear Research (CERN), Geneva, Switzerland\\
% ...
$ ^{43}$Nikhef National Institute for Subatomic Physics, Amsterdam, The Netherlands\\
% ...
$ ^{48}$H.H. Wills Physics Laboratory, University of Bristol, Bristol, United Kingdom\\
% ...
$ ^{53}$School of Physics and Astronomy, University of Glasgow, Glasgow, United Kingdom\\
% ...
$ ^{59}$University of Cincinnati, Cincinnati, OH, United States\\
% ...
}
\end{flushleft}